\documentclass[useAMS,usenatbib]{mn2e}
\usepackage{graphicx}

\def\kpc{\,\rm kpc}
\def\msun{M_\odot}
\def\fs{f_{\rm S}}
\def\fsi{f_{\rm S_I}}
\def\I0{I_{\rm 0}}
\def\tE{t_{\rm E}}
\def\te{t_{\rm E}}
\def\t0{t_{\rm 0}}
\def\u0{u_{\rm 0}}

\newcommand{\eg}{{e.g.},\,}
\newcommand{\ie}{{i.e.},\,}
\newcommand{\apj}{{Astrophysical Journal}}
\newcommand{\apjl}{{Astrophysical Journal Letters}}
\newcommand{\apjs}{{Astrophysical Journal Supplement Series}}
\newcommand{\araa}{{ARA\&A}}
\newcommand{\aap}{{Astronomy \& Astrophysics}}
\newcommand{\mnras}{{MNRAS}}
\newcommand{\aj}{{\apj}}

\title[The OGLE View of Microlensing towards the Magellanic Clouds. I. OGLE--II LMC]
{The OGLE View of Microlensing towards the Magellanic Clouds. I. A Trickle of Events in the OGLE--II LMC data.  \thanks{Based on
    observations obtained with the 1.3~m Warsaw telescope at the Las Campanas Observatory of the Carnegie Institution of Washington.}}
\author[{\L}. Wyrzykowski et al.]
{{\L}. Wyrzykowski$^{1,2}$\thanks{email: wyrzykow@ast.cam.ac.uk, name
    pronunciation: {\it Woocash Vizhikovsky}}, S. Koz{\l}owski$^3$,
  J. Skowron$^2$, V. Belokurov$^1$, M. C. Smith$^1$, \newauthor
  A. Udalski$^2$, M. K. Szyma{\'n}ski$^2$, M. Kubiak$^2$,
  G. Pietrzy{\'n}ski$^{2,4}$,
  \newauthor I. Soszy{\'n}ski$^2$, O. Szewczyk$^{2,4}$ and K. {\.Z}ebru{\'n}$^2$\\
  $^1$ Institute of Astronomy, University of Cambridge,
  Madingley~Road,
  Cambridge~CB3~0HA,~UK \\
  $^2$ Warsaw University Astronomical Observatory, Al.~Ujazdowskie~4, 00-478~Warszawa, Poland \\
  $^3$ Department of Astronomy, Ohio State University, 140 W. 18th Ave., Columbus, OH 43210, USA\\
  $^4$ Universidad de Concepci{\'o}n, Departamento de Fisica,
  Astronomy Group, Casilla 160-C, Concepci{\'o}n, Chile\\ }

\begin{document}

\date{Accepted  ... Received  ...}

\pagerange{\pageref{firstpage}--\pageref{lastpage}} \pubyear{2009}

\maketitle

\label{firstpage}

\begin{abstract}

We present the results from the OGLE--II survey (1996-2000) towards the Large
Magellanic Cloud (LMC), which has the aim of detecting the microlensing
phenomena caused by dark matter compact objects in the Galactic Halo (Machos).

We use high resolution HST images of the OGLE fields and 
derive the correction for the number of monitored stars in each field.  
This also yield blending distributions which we use in
'catalogue level' Monte Carlo simulations of the microlensing events
in order to calculate the detection efficiency of the events.

We detect two candidates for microlensing events in the All Stars
Sample, which translates into an optical depth of $0.43\pm0.33\times
10^{-7}$.  If both events were due to Macho the fraction
of mass of compact dark matter objects in the Galactic halo would be
$8 \pm 6$ per cent.  This optical depth, however, along with the
characteristics of the events,  seems to be consistent with the
self--lensing scenario, i.e., self--lensing alone is sufficient to
explain the observed microlensing signal. Our results indicate a
non-detection of Machos lensing towards the LMC with an upper limit on
their abundance in the Galactic halo of 19 per cent for $M=0.4\msun$
and 10 per cent for masses between 0.01 and 0.2 $\msun$.

\end{abstract}

\section{Introduction}

Although we are constantly learning new things about our Galaxy, some
major questions about its structure still remain unanswered.  Among
them is the composition of the Galaxy halo.  Solving this mystery has
been the main goal of the large photometric surveys launched in the 1990s.
The proposal of \citet{Paczynski1986} has been realised by the surveys
like MACHO \citep{MACHO}, OGLE \citep{Udalski1993}, EROS \citep{EROS}
and MOA \citep{MOA}.  These groups continuously monitored millions of
stars in both Magellanic Clouds in order to detect the transient
amplification of brightness caused by objects crossing the observer--source
line of sight in a phenomenon called gravitational microlensing
\citep{Paczynski1996}. The rate of such events depends on the number, mass
and kinematics of objects populating the line of sight, in the disk
and the halo of our Galaxy and in the Clouds. The main hypothesis all
the surveys have been testing is whether the halo of our Galaxy
contains dark matter in the form of the MAssive Compact Halo Objects,
called Machos.

If the Galactic halo were entirely made of Machos, with masses between
$10^{-8}$ and $10^8 \msun$, in a project like OGLE--II we should detect about five
microlensing events per year towards the Large Magellanic Cloud (LMC),
and the corresponding optical depth should be $\tau_{\rm Macho} \approx
4.7 \times 10^{-7}$ (\citealt{BennettMACHOLMC}). However, the actual
optical depth measurements reported by the microlensing experiments
are significantly smaller.

The MACHO collaboration presented a sample of 10
microlensing event candidates towards the LMC and concluded that the
fraction of dark matter in Machos $f = M_{\rm Macho}$/$M_{\rm
  halo} \approx$ 20 per cent, suggesting the existence of a new,
previously unknown population of objects with masses $\sim 0.5 \msun$
(\citealt{AlcockMACHOLMC}, \citealt{BennettMACHOLMC}).  Their estimate
of the optical depth towards the LMC is $\tau_{\rm LMC} = (1.0 \pm
0.3) \times 10^{-7}$.  However, the EROS group reported a null result
and derived an upper limit on the optical depth due to Machos of
$\tau_{\rm Macho} < 0.36 \times 10^{-7}$, which implies $f < 8$ per cent
(\citealt{TisserandEROSLMC}).  
MACHO's sample of events is likely to have some contamination with
variable stars, novae, and supernovae \citep{Belokurov2003}, and
additional data indeed shows the presence of small contamination \citep{BennettBeckerTomaney2005}. 
However, the optical depth we quote from MACHO has been
corrected for this contamination \citep{BennettMACHOLMC}.
The large discrepancy between the results may be due 
to the different samples of stars used by these groups or the fact that they probe different regions of the LMC. 
However, it is unclear whether these fully explain the discrepancy.

Independent searches for Machos have also been carried out towards
the Andromeda Galaxy (M31) using pixel-lensing method.  For example,
\citet{CalchiNovati2005} found 6 events and estimated the Machos halo
mass fraction to be higher than 20 per cent.  On the other hand,
\citet{BelokurovM31} presented an automatic procedure that selects
unambiguous microlensing events in a pixel light curve data set with a
total of only 3 ``gold-plated'' candidates, thus emphasizing the lack
of large numbers of clear-cut microlensing events towards the M31.
\citet{deJong2006} showed 14 events towards M31 and noticed that the observed event rate is consistent with the self-lensing predictions, however their results were still inconclusive, mainly due to low statistics.

In this paper we present the results of the search for microlensing
events towards the Large Magellanic Cloud in the independent data set
gathered by the Optical Gravitational Lensing Experiment (OGLE) in
its second phase during the years from 1996 to 2000.  The paper is
organized as follows.  First, the observational data are described along
with the error bar correction procedure.  Next, the event search
algorithm and its results are presented.  It is followed by the
description of the analysis of blending and its impact on the
optical depth measurement and the event's detection efficiency.
Finally, the optical depth is calculated and the results are
discussed.

\section{Observational Data}

The data used in this paper were collected during the second phase of
the OGLE survey, in the years 1996 -- 2000 towards the Large Magellanic
Cloud (LMC).  The OGLE project used its own dedicated Warsaw Telescope
in the Las Campanas Observatory, Chile.  Details of the
instrumentation for OGLE--II can be found in \citet{OGLE2}.

\begin{figure*}
\includegraphics[width=12cm]{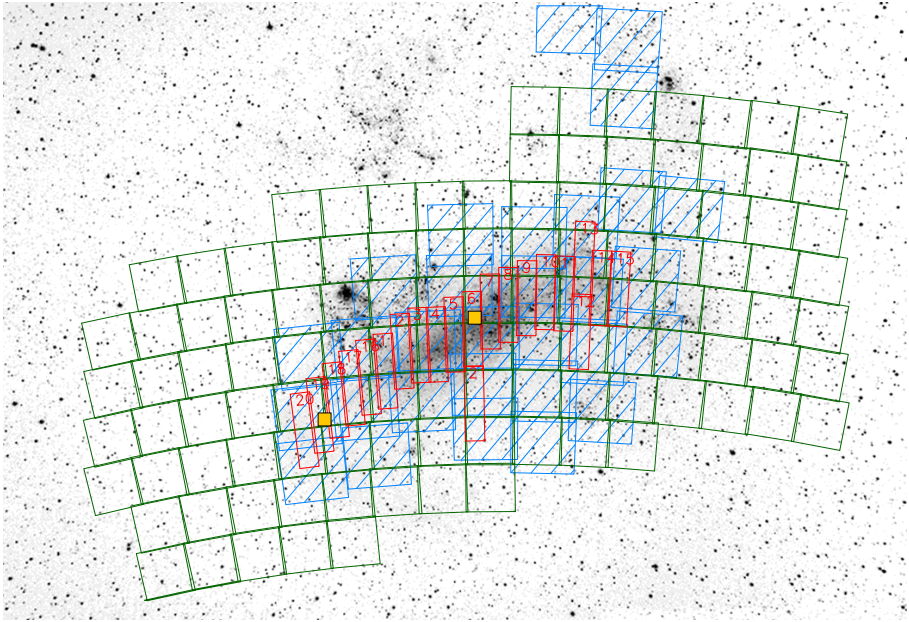}
\caption{Positions of the OGLE--II LMC fields (red rectangles with
  labels). Also shown are all OGLE--III (green squares) and MACHO project
  fields (blue dashed squares). EROS fields are not shown here as they cover
  nearly the whole picture. The two small gold-filled squares show
  the positions of the HST fields used for our blending determination.
  Background image credit: Peter Ward.}
\label{fig:fields}
 \end{figure*}

In brief, a 2k$\times$2k CCD with a pixel size of 0.417~arcsec was
operated in the drift-scan mode, giving an actual size for each observed field
of 2k$\times$8k.  There were 21 fields observed towards the LMC
covering a total of 4.72 square degrees. Their locations are shown in
Fig. \ref{fig:fields}.  Fields are listed in Table \ref{tab:fields}, which 
contains the coordinates of centers of the fields, the number of good objects in
the $I-$band, the blending-corrected number of stars (see Section
\ref{sec:blending}) and the blending density group. By 'good', we mean all
objects having at least 80 epochs during the entire time span of the
OGLE--II (from about 15 to 25 per cent of all collected frames of a field) 
and mean magnitude brighter than $20.4$ mag. The limiting magnitude
was chosen to be at the peak of the observed luminosity function.

The images in the $I$ and $V$ bands were de-biased and flat-field
corrected ``on-the-fly''.  The photometric pipeline was based on the
Difference Image Analysis method (DIA, \citealt{WozniakDIA,
  AlardLuptonDIA}).  Photometric databases were created for both
pass-bands as described in \citet{Szymanski2005} and are available on-line\footnote{http://ogledb.astrouw.edu.pl}.

The OGLE--II observations of the LMC began in December 1996
(HJD=2450446), but during the first year only the very central fields
(i.e. those that are, in general, denser) were observed (fields 1--10
and 12).  In November 1997 (HJD=2450726), fields 11 and 13--20 were
added to the observing queue, and the addition of the field LMC\_SC21
in January 1998 (HJD=2450831) eventually formed the 
entire set of OGLE--II fields.  All 21 fields were monitored
continuously until November 2000 (HJD=2451874), yielding about 500 and
300 frames per field in the $I$ band for dense and sparse fields,
respectively.  In the $V$ band, there were no more than 50 frames per
field collected in total during the duration of the project.  On
average, each field was observed every third night in the $I$ band, and
every 11-th night in the $V$ band.  Mean seeing was 1.35 arcsec in the $I$ and
1.37 arcsec in $V$. Template images were created by stacking the best
seeing images, resulting in images with seeing of about 1.1 arcsec.
The template images contained about 5.5 million objects suitable for our study in the $I$ and $V$ band.

The detection of microlensing events and the efficiency
determination have been performed using $I-$ band data only, as these
were far more numerous and were sampled more frequently as compared to
$V-$band photometry.  

In this study we also occasionally made use of OGLE--III LMC data from years 2001--2008, which
covered a much wider area and included the OGLE--II fields.  Details of
the instrumentation and photometric pipeline can be found in
\citet{Udalski2003}.  Analysis of OGLE--III LMC data will be presented
separately in a forthcoming paper (Wyrzykowski et al., in
preparation).

Additionally, for the two detected events, we also use the data
gathered by the MACHO group and provided on their WWW
interface\footnote{http://wwwmacho.mcmaster.ca/Data/MachoData.html}. The
reductions of the images and extraction of the photometry were
performed with Difference Image Analysis following
\citet{Kozlowski2007}.

\begin{table*}
\caption{The OGLE--II LMC fields.}
\label{tab:fields}
\centering
\begin{tabular}{cccrrcc}
\hline
\noalign{\vskip5pt}
Field & $RA_{J2000}$ & $Dec_{J2000}$ & \multicolumn{2}{c}{No of stars} & Stellar density & Density level\\
& & & template & estimated real & [stars/sq.arcmin] & \\
\noalign{\vskip5pt}
\hline
\noalign{\vskip5pt}
LMC\_SC1 & 5:33:48.84 & -70:06:07.9 & 288029 & 668799 & 367 & dense \\
LMC\_SC2 & 5:31:17.17 & -69:51:57.0 & 332267 & 771301 & 424 & dense \\
LMC\_SC3 & 5:28:47.95 & -69:48:05.6 & 355075 & 822913 & 453 & dense \\
LMC\_SC4 & 5:26:17.82 & -69:48:06.1 & 390484 & 904432 & 498 & dense \\
LMC\_SC5 & 5:23:47.79 & -69:41:03.3 & 379709 & 878472 & 484 & dense \\
LMC\_SC6 & 5:21:18.34 & -69:37:10.3 & 392781 & 909775 & 501 & dense \\
LMC\_SC7 & 5:18:48.01 & -69:24:09.9 & 388918 & 900339 & 496 & dense \\
LMC\_SC8 & 5:16:17.91 & -69:19:13.7 & 349202 & 807890 & 445 & dense \\
LMC\_SC9 & 5:13:47.92 & -69:14:05.8 & 289391 & 671459 & 369 & dense \\
LMC\_SC10 & 5:11:16.10 & -69:09:17.3 & 258706 & 599375 & 330 & dense \\
LMC\_SC11 & 5:08:41.06 & -69:10:03.1 & 253395 & 588432 & 323 & dense \\
LMC\_SC12 & 5:06:16.15 & -69:38:18.3 & 186951 & 318893 & 238 & sparse \\
LMC\_SC13 & 5:06:14.18 & -68:43:30.3 & 214229 & 365917 & 273 & sparse \\
LMC\_SC14 & 5:03:48.62 & -69:04:44.3 & 222569 & 380394 & 284 & sparse \\
LMC\_SC15 & 5:01:17.27 & -69:04:42.9 & 170657 & 291859 & 218 & sparse \\
LMC\_SC16 & 5:36:17.62 & -70:09:41.9 & 247981 & 575953 & 316 & dense \\
LMC\_SC17 & 5:38:47.47 & -70:16:44.1 & 194661 & 332187 & 248 & sparse \\
LMC\_SC18 & 5:41:17.77 & -70:24:48.0 & 166998 & 285126 & 213 & sparse \\
LMC\_SC19 & 5:43:47.91 & -70:34:42.7 & 156353 & 267126 & 199 & sparse \\
LMC\_SC20 & 5:46:17.85 & -70:44:50.0 & 152186 & 260515 & 194 & sparse \\
LMC\_SC21 & 5:21:13.73 & -70:33:18.7 & 143979 & 244667 & 184 & sparse \\
\hline
total & \multicolumn{2}{c}{} & 5534521 & 11845824 & \\
\noalign{\vskip5pt}
\hline
\end{tabular}


\medskip
\begin{flushleft}
Coordinates point to the centre of the field, each being $14' \times
56'$. The number of 'good' objects in the template is provided (number of data points $>80$ and
 mean $I$-band magnitude $< 20.4$ mag) together with the estimated number of
real monitored stars (see Section \ref{sec:blending}). Stellar density
in number of stars per square arc minute was used to classify fields
into {\it dense} or {\it sparse} classes with the threshold of 300
stars/sq.arcmin.
\end{flushleft}
\end{table*}

\section{Error-bars correction}
\label{sec:errors}

Photometric error bars calculated using the Difference Image Analysis
method are known to be generally underestimated.  This effect is most
pronounced for bright stars, as noted already by the authors of the DIA
method \citep{AlardLuptonDIA}.  In microlensing studies this affects
the brightest parts of events, increasing the $\chi^2$ values in
microlensing model fitting.  
In order to increase the detectability of events, we derived and
applied an error bar correction factor.

Typically, error bars in microlensing surveys are corrected by
calculating one single correction factor for an entire light curve,
derived at the event's baseline (assuming the baseline is
constant). This approach, however, does not account for the 
need for varying correction factors for different apparent magnitudes, which
in case of highly magnified events leaves the error bars at the peak of
the event significantly underestimated. Also, as shown in
\citet{varbaseline}, some events exhibit intrinsic variability in
their baselines due to variability of the microlensed source or one of
the blended stars, in which case the derived photometric error bars
are overestimated.

\begin{figure}
\center
 \includegraphics[width=6.5cm]{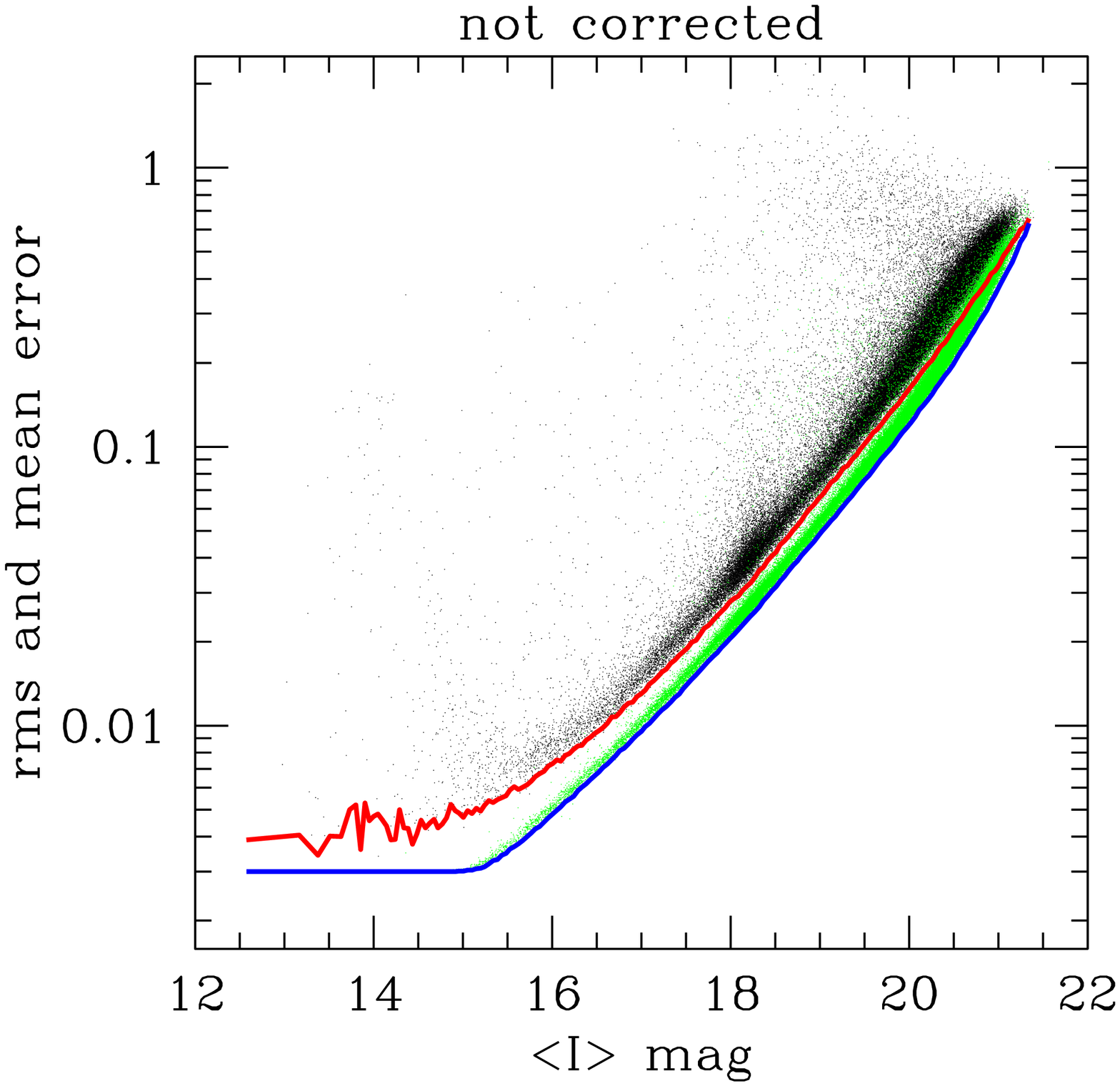}
  \includegraphics[width=6.5cm]{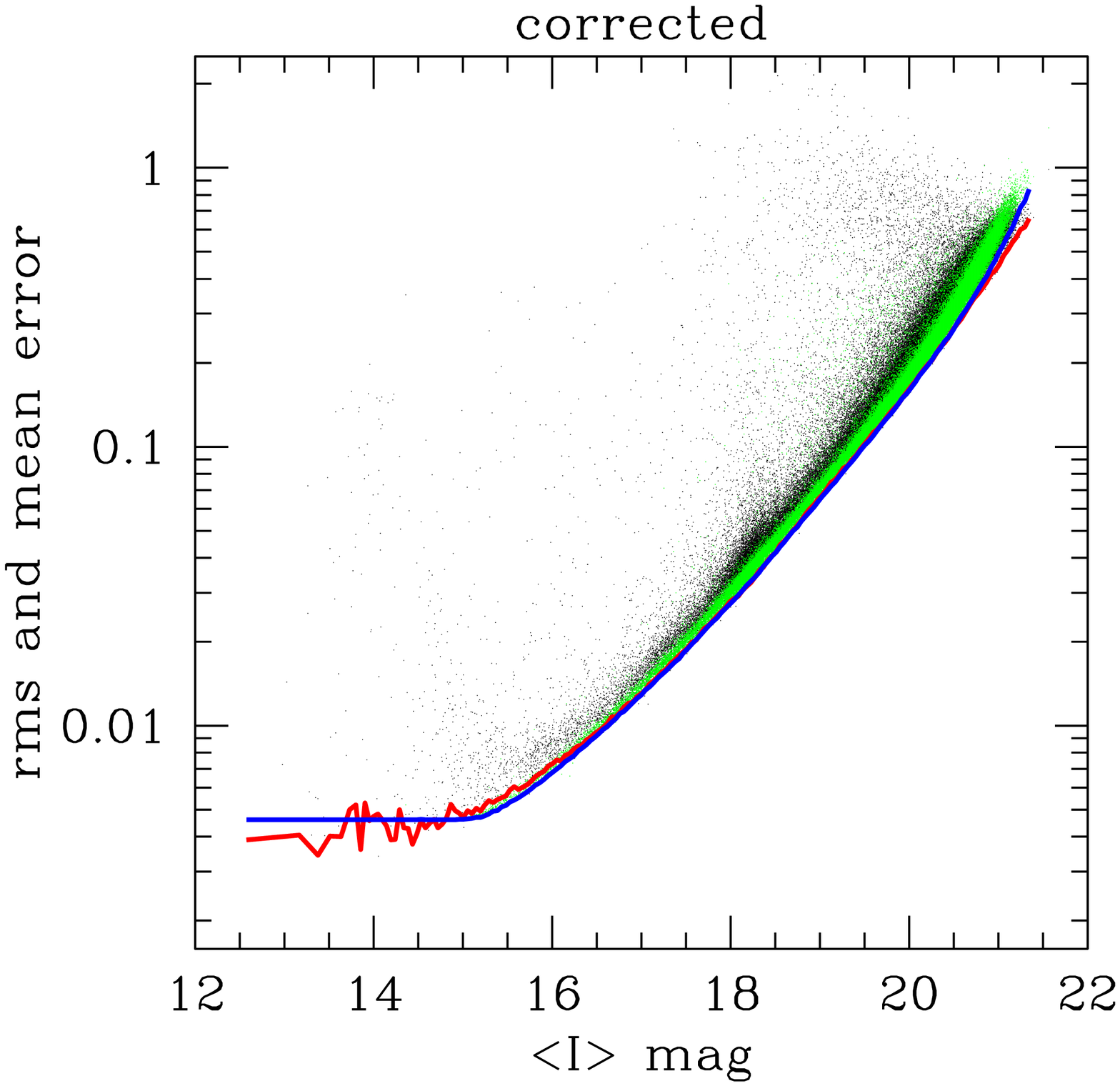}
\caption{Mean error (green) and {\it rms} (black) for a sample 
of stars from OGLE--II  LMC\_SC1 field before (upper panel) and after (lower panel) the correction of error bars. 
Lines indicate the lower outlines of the {\it rms} (red) and mean error bar (blue).}
\label{fig:errors}
\end{figure}

In this study, we processed the entire OGLE--II data set and determined
the correction coefficients in an empirical way.  For each OGLE--II
field, we extracted the $I$-band light curves of stars and compared
their intrinsic error-weighted {\it rms} with the mean error returned
by the photometry pipeline.  In order to produce statistically
comparable results, in each light curve we selected only the first 250
measurements for further analysis, as different fields had different
numbers of frames taken.  In the plot of {\it rms} vs magnitude (see
Fig. \ref{fig:errors}), the lowest boundary of the {\it rms} is
occupied by the least variable (\ie constant) stars.  On the
other hand, the lowest outline of mean error bar is defined by the
errors assigned to the best measurements at given magnitude.  In an
ideal situation these two curves should follow each other, \ie every
constant star should have its {\it rms} equal to its mean error bar.
As can be seen from the upper panel of Fig. \ref{fig:errors}, for our data
the mean errors were systematically lower than the {\it rms}.

\begin{table}
\centering
\caption{Error correction coefficients for OGLE--II LMC fields.}
\label{tab:errorcor}
\begin{tabular}{ccc}
\hline
Field & $\gamma$ & $\epsilon$ \\
\hline
LMC\_SC1 & 1.327934 & 0.002309 \\
LMC\_SC2 & 1.534320 & 0.000000 \\
LMC\_SC3 & 1.514598 & 0.000000 \\
LMC\_SC4 & 1.534963 & 0.000000 \\
LMC\_SC5 & 1.552337 & 0.000000 \\
LMC\_SC6 & 1.512235 & 0.000000 \\
LMC\_SC7 & 1.546070 & 0.000000 \\
LMC\_SC8 & 1.376516 & 0.001565 \\
LMC\_SC9 & 1.432857 & 0.001066 \\
LMC\_SC10 & 1.306555 & 0.001539 \\
LMC\_SC11 & 1.084232 & 0.002695 \\
LMC\_SC12 & 1.325315 & 0.001657 \\
LMC\_SC13 & 1.369874 & 0.001256 \\
LMC\_SC14 & 0.932450 & 0.002931 \\
LMC\_SC15 & 1.168966 & 0.002046 \\
LMC\_SC16 & 1.151916 & 0.002628 \\
LMC\_SC17 & 1.149929 & 0.002365 \\
LMC\_SC18 & 1.205946 & 0.002624 \\
LMC\_SC19 & 1.128117 & 0.003138 \\
LMC\_SC20 & 1.203747 & 0.002623 \\
LMC\_SC21 & 1.301942 & 0.002467 \\
\hline
\end{tabular}
\end{table}

In order to account for this systematic shift we have fitted the
following two-parameter formula which minimised the difference between
lower outlines of the {\it rms} and mean error bars:
\begin{equation}
\Delta I_{cor} = \sqrt{ ( \gamma \Delta I )^2 + \epsilon^2 },
\label{eq:errors}
\end{equation}
where $\Delta I$ is the error bar returned by the photometry pipeline,
$\gamma$ and $\epsilon$ are the correction coefficients and $\Delta
I_{cor}$ is the corrected error bar.

The lower panel of Fig.\ref{fig:errors} shows the {\it rms} and mean
error bars after applying the corrections to each light curve's
individual error bar measurements using Equation \ref{eq:errors}.

Table \ref{tab:errorcor} presents $\gamma$ and $\epsilon$ coefficients
derived for each OGLE--II LMC field.  
The mean $\gamma$ and $\epsilon$ for
all LMC fields were $1.32$ and $0.0016$, respectively.
Derived correction coefficients are suitable for use in any study involving OGLE--II Large Magellanic Cloud photometric data.

As a by-product of the error bar analysis we derived the formula for
scaling real error bars from some reference light curve to any
magnitude, which is necessary for the simulation of light
curves:
\begin{eqnarray}
\Delta I_{\rm sim} = \Delta \, I_{\rm ref} 10^{0.35 ( I_{\rm sim} - I_{\rm ref} ) }, & {\rm for}~I_{\rm sim} \ge 15.0 & \\
\nonumber
\Delta I_{\rm sim} = 0.003, & {\rm for}~I_{\rm sim} < 15.0, & 
\label{eq:simerr}
\end{eqnarray}
where $I_{\rm sim}$ is the simulated magnitude for which the error bar ($\Delta
I_{\rm sim}$) is required, $I_{\rm ref}$ and $\Delta I_{\rm ref}$ are the
magnitude and the error bar of the reference star at a given epoch.
These error bars still need to be corrected using eq. (\ref{eq:errors}).

\section{Search procedure}
\label{sec:search}

\begin{table*}
\centering
\caption{Selection criteria for microlensing events in the OGLE--II data}
\label{tab:conditions}
\begin{tabular}[h]{c|lrr}
\hline
Cut no. & & & No. of objects left \\
\hline     
0 & Selection of ``good'' objects & $N>80$, $\langle I \rangle \le 20.4$ mag  &  5,534,521\\
& & \\

1 & Significant bump over baseline & $\displaystyle\sum_{peak} \sigma_i > 30.0 $ & 9,390 \\
& & \\

2 & ``Bumper'' cut$^\dagger$ & $\langle I \rangle>19.0~mag $,  $\langle V-I \rangle>0.5~mag$ & 7,867 \\ 
& &  &\\

3 & Microlensing fit better than constant line fit & ${{\chi^2_{line}-\chi^2_{{\mu} 4}}\over{{\chi^2_{{\mu}4}\over{N_{dof,\mu4}}}\sqrt{2N_{dof,\mu4,peak}}}} > 140$  & 1,020 \\
& & & \\

4 & Number of points at the peak$^{\ast}$ & $N_{peak} > 6$  & 968 \\ 
& & & \\

5 & Microlensing fit better than supernova fit, & $\chi^2_{SN} > MIN(\chi^2,\chi^2_{{\mu}4})$ & 607 \\
& & & \\

6 & Peak within the data span   & $446 \le {\t0} \le 1874$   & 591 \\
& [HJD-2450000]    &         &         \\
& &  & \\

7 & Blended fit converged & $0< \fs < 1.2$ & 113 \\
& & & \\

8 & Conditions on goodness of microlensing fit &  ${{\chi^2}\over{N_{dof}}} \le 2.3 $ and ${{\chi_{\mu4,peak}^2}\over{N_{dof,\mu4,peak}}} \le 5 $ & 4 \\
& (global and at the peak) & & \\

9 & Time-scale cut & $1 \le {\tE} \le 500$ & 2 \\
& $[d]$ &  & \\
&  & & \\

10 & Impact parameter cut &  $0 < {\u0} \le 1 $ & 2 \\
& & & \\

\hline
\end{tabular}
\\
$^{\dagger}$ magnitudes as in the field LMC\_SC1 (shifted according to the position of the center of Red Clump) \\
$^{\ast}$in the range of $\t0 \pm t_{\rm E\mu 4}$ \\
\end{table*}

\begin{figure}
\center
\includegraphics[width=5.5cm]{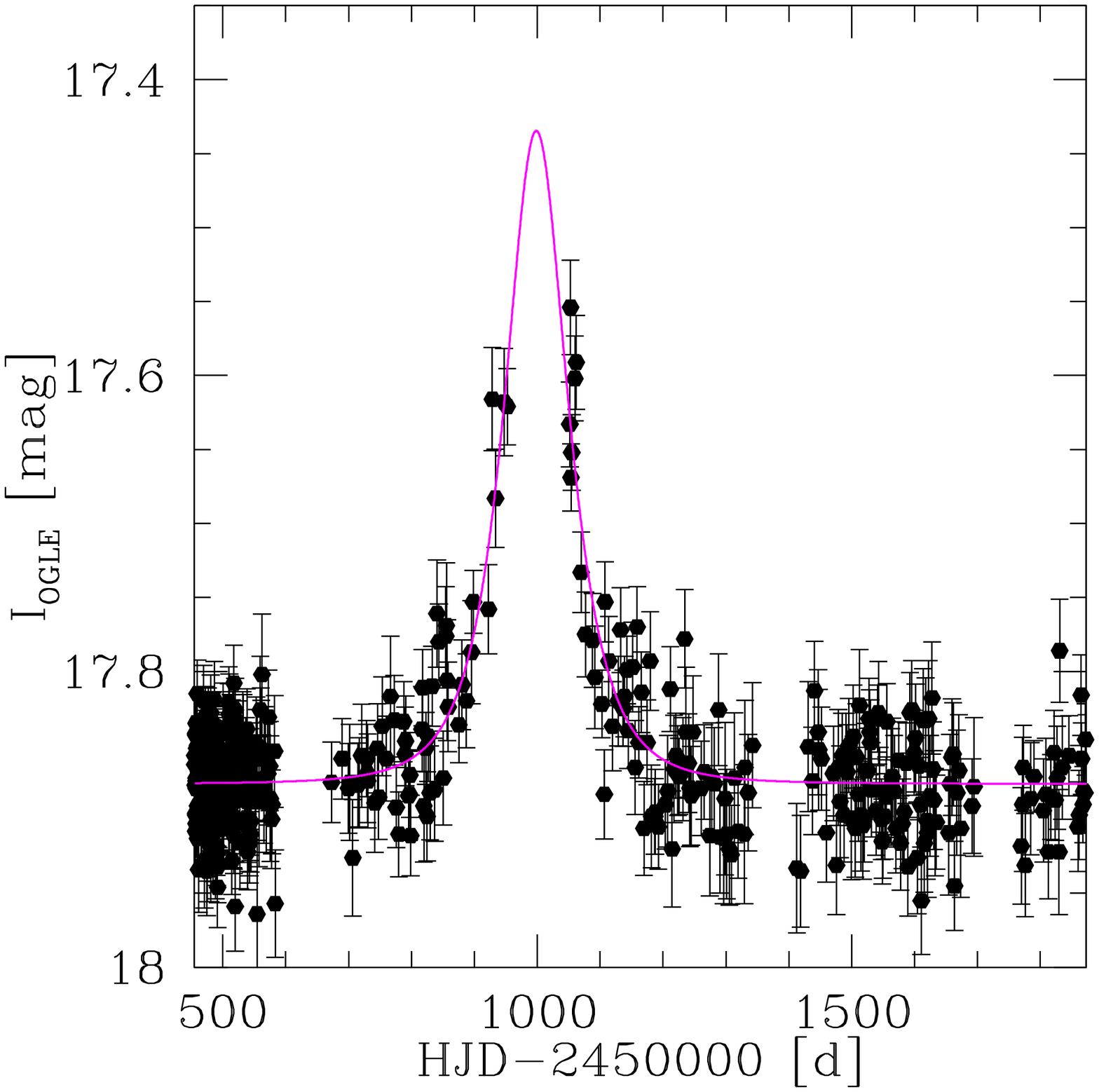}
\includegraphics[width=5.5cm]{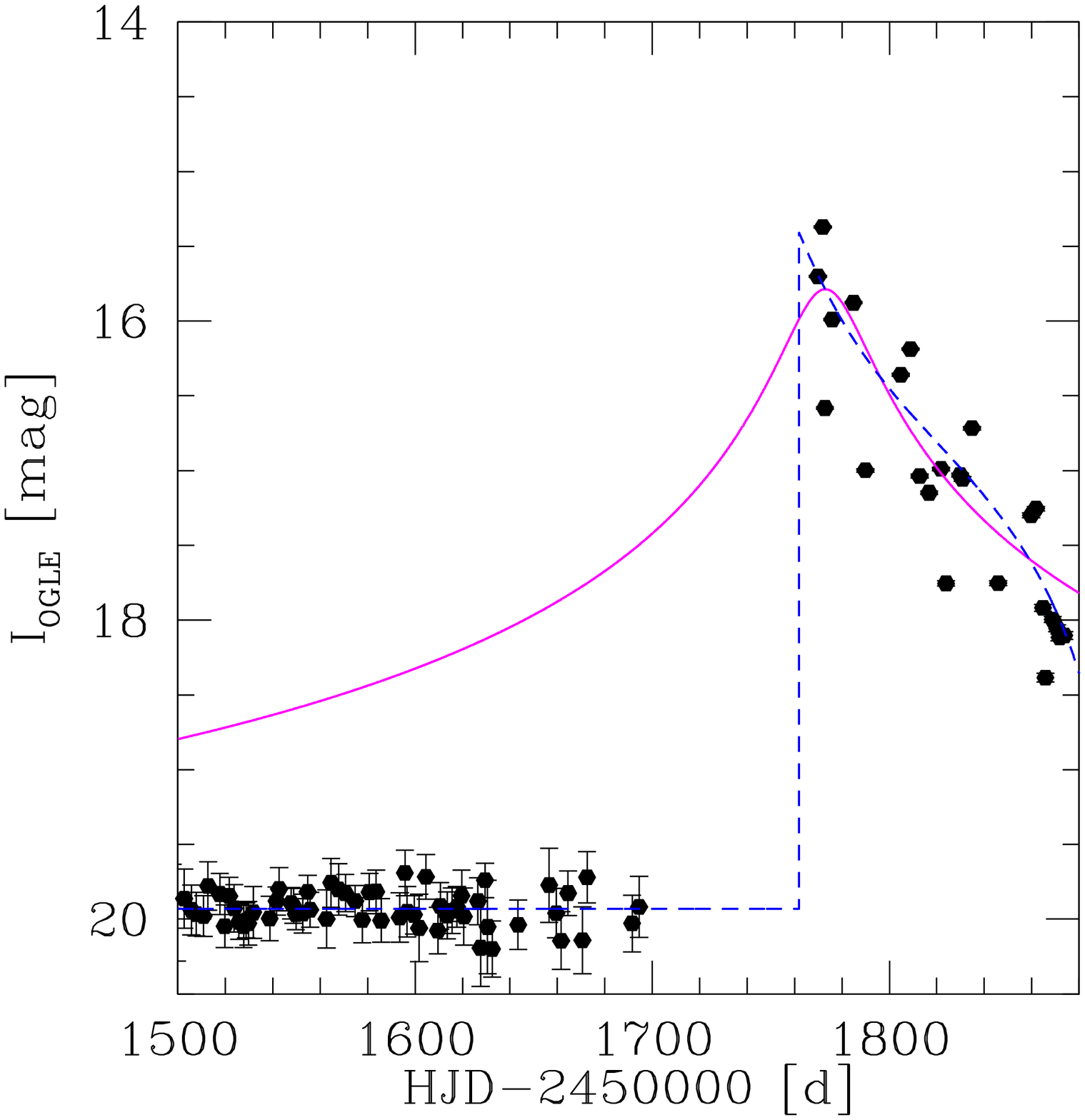}
\caption{Two sample light-curves for false--positive events, which
  were rejected in the search procedure. Top: ``blue bumper'' (which exhibited a
  second bump in the OGLLE--III data). Bottom: possible nova or supernova with additional variability in the declining part. Solid magenta line is the best microlensing model, dashed blue line is a nova model.  }
\label{fig:badevents}
\end{figure}

We searched for microlensing events among 5.5 million objects, selected from the
database as having more than 80 data points and mean magnitude in $I$ band
brighter than 20.4 mag (cut 0). This defined our All Stars Sample. 
All light curves now have their measurements' error bars corrected as described in Section \ref{sec:errors}.

Table \ref{tab:conditions} presents all of the cuts, along with
the number of objects left after each cut.  All parameters for this
procedure were derived and fine-tuned for OGLE--II LMC data after
carrying-out Monte Carlo simulations of microlensing events, using the
simulation procedures described in Section \ref{sec:eff}.

The first step of the search procedure is similar to the one used in
the search for microlensing events in the OGLE--II Galactic bulge data
\citep{SumiOGLEbulge}.  All objects were searched for a bump over a
baseline, allowing for some variability in the baseline.

For a given light curve, we introduce an outer window with width equal to half of the total observational time-span.
The significance of each data point $\sigma_i$ with respect to the outer window was calculated via
\begin{equation} \label{eq:signif}
\sigma_i = { I_{med,w} - I_i \over \sqrt{{\Delta I_i}^2 + {\sigma_{w}}^2 }}
\end{equation}
where $I_i$ is the magnitude of the point {\it i} and $\Delta I_i$ is
its error bar, while $I_{med,w}$ and $\sigma_{w}$ are the median and the
{\it rms} in the outer window, respectively.  The bump was
detected when there were at least 4 consecutive data points above
a $\sigma>1.4$ threshold and 3 of them where above a $\sigma>2.5$
threshold. Up to 2 peaks were allowed in one light curve, but then
only the one with the highest $\sigma_{max}$ was chosen.  If the sum of $\sigma$s over the peak data points was higher than 30.0 the light curve was selected (cut 1).

In cut 2, we masked out possible ``blue bumpers''
\citep{AlcockMACHOLMC} -- variable stars located on the bright blue
part of the Main Sequence, exhibiting outbursts that can look similar
to the microlensing phenomenon.  The upper panel of Fig. \ref{fig:badevents} 
shows an example of a ``blue bumper'' star,
rejected due to its magnitude and colour. 
This particular object also exhibited additional significant variation in the OGLE--III data in subsequent years.
 
In the next step, the remaining light curves were fitted with the
standard \citet{Paczynski1996} microlensing model, (\ie
a point-source -- point-lens microlensing event) which is described as:

\begin{equation}
\label{eq:I}
I = \I0 - 2.5\log\left[ \fs A + (1-\fs)\right],
\end{equation}

\noindent
where

\begin{equation}
\label{eq:A}
A= { u^2 + 2 \over u\sqrt{u^2+4} } \qquad {\rm and} \qquad u= \sqrt{\u0^2 + {{(t-\t0)^2} \over {\tE^2}}}.
\end{equation}

The fitted parameters are: $\t0$ -- the time of the maximum of the peak,
$\tE$ -- the Einstein radius crossing time (event's time-scale), $\u0$ --
the event's impact parameter, $\I0$ -- the baseline magnitude in the $I$ band and
$\fs$ -- the blending fraction (ratio of lensed source flux to total blends' flux in the $I$ band).
The fits were performed in two ways, namely with blending parameter fixed
$\fs=1$, \ie with no blending, and with $\fs$ being free. 
For clarity, parameters of the non-blended (4 parameters) model are
given the subscript $\mu 4$. 

Then we compared the 4-parameter microlensing model (with blending
fixed to 1 to assure convergence) with a constant line fit (cut 3). 
A similar cut was also applied in the EROS analysis \citep{TisserandEROSLMC},
where it proved to be very efficient.
$N_{dof,\mu 4}$ and $N_{dof,\mu 4,peak}$ are the degrees of freedom of the fit calculated using numbers of data points in the entire light curve and around the peak, respectively.
This cut removed many low signal-to-noise bumps and allowed us to
select only the outstanding bumps.

In cut 4, we required that there were more than 6 data points
around the microlensing peak in the range of $\t0 \pm t_{\rm E\mu 4}$.
This meant that we had enough data points in the bump to analyse.

Next we attempted to remove all plausible novae, supernovae and all
kinds of asymmetric bumps contaminating our sample (cut 5).  We
fitted the light curves with an general asymmetric model, 
following the formula used for the novae search in e.g.
\citet{Feeney2005}, which comprises of two scalable exponents and a
constant baseline.  We compared its goodness-of-fit with
the $\chi^2$ of microlensing model, taking the smaller of
blended and non-blended fits.
This cut was sensitive to all sorts of asymmetric light curves.  Among
the removed objects were several dwarf novae candidates, as well as
long-term varying objects and several eruptive Be stars. These Be
stars were fainter or redder than our Blue Bumper cut and typically
exhibited a flat baseline, a sharp rise and a slow decline (\eg
\citealt{Keller2002}).

Assuming a supernova rate of 0.5 SNe yr$^{-1}$ deg$^{-2}$ above $V=20$ mag
\citep{AlcockMACHOLMC}, then for our experiment of 4 years covering 4.5
square degrees and assuming detection efficiency of about 20 per cent we
should expect about 1 supernova in our data set.  Visual inspection of
the light curves rejected at the supernova fit stage (cut 5) indeed
showed one plausible candidate for a supernova (see lower plot of Figure
\ref{fig:badevents}).

In the following step (cut 6), we removed all light curves in which the
peak occurred before the beginning or after the end of the time-span
of the OGLE--II observations.

Next, in cut 7 we selected only those candidates whose blended
microlensing fit has converged with blending parameter $\fs<1.2$.
We allowed for small amount of 'negative' blending following previous
studies (\eg \citealt{Park2004}, \citealt{Smith2007blending}) showing regular event can be
affected by this effect.  The vast majority of the light curves
removed in cut 7 had $\fs$ significantly larger that 2, indicating
that the blending fit did not converge.

In cut 8, we requested the reduced $\chi^2$ of the blended
microlensing fit was less than 2.3 and the $\chi^2$ of the non-blended
model calculated on the peak only (within 1 $t_{\rm E\mu4}$ around
$\t0$) was less than 5.0.  This stage removed obvious variable stars
with a single bump lying above the rest of the light curve to
which the microlensing model was fit quite poorly.  We were able to
apply this cut uniformly to all light curves, independently on their
magnitude, thanks to the fact that the error bars were corrected and any
dependence on the measurement brightness was accounted for (Section
\ref{sec:errors}).

The final cuts 9 and 10 are applied to select candidates suitable
for the optical depth determination.  For this, we used microlensing
parameters obtained with the fitting procedure, and required
the time-scales $\tE$ to be between 1 and 500 days and
impact parameter $\u0$ less than 1 Einstein radius.


\begin{table*}
\caption{Microlensing candidates detected in the OGLE--II LMC data.}
\label{tab:events}
\begin{center}
\begin{tabular}{ccccccccc}
\hline
Event's name             & RA          & Dec       & field & database      &  baseline $I$      & baseline $V$ & source $I$ & source $V-I$\\
                               & [J2000.0]   & [J2000.0] &         & star id & [mag]                   & [mag]                          & [mag] &          [mag]\\
\hline
OGLE-LMC-01          & 5:16:53.26  & -69:16:30.1 &  LMC\_SC8 & 235928 & 19.91    & 20.65          & 19.90            & 0.887$^*$ \\
{\scriptsize(EWS: OGLE-1999-LMC-01)} &             &          &      &     	 & $\pm$0.01 &$\pm$0.14     &	$\pm$0.06   & $\pm$0.007  \\
{\scriptsize(MACHO-99-LMC-2)}  & & & & & & & \\

 & & & & & & & \\
OGLE-LMC-02          & 5:30:48.00  & -69:54:33.6 & LMC\_SC2 & 170259 & 20.42     & 20.68  & 20.53 &  0.46 \\
       &     &                        &             &   							&$\pm$0.02    & $\pm$0.13 & $\pm0.38$ & $\pm$0.03 \\
\hline			       

\end{tabular}
\\
$^*$ colour derived using MACHO $B$-band data\\
\end{center}
\end{table*}


\begin{figure*}
\center
\includegraphics[width=8.5cm]{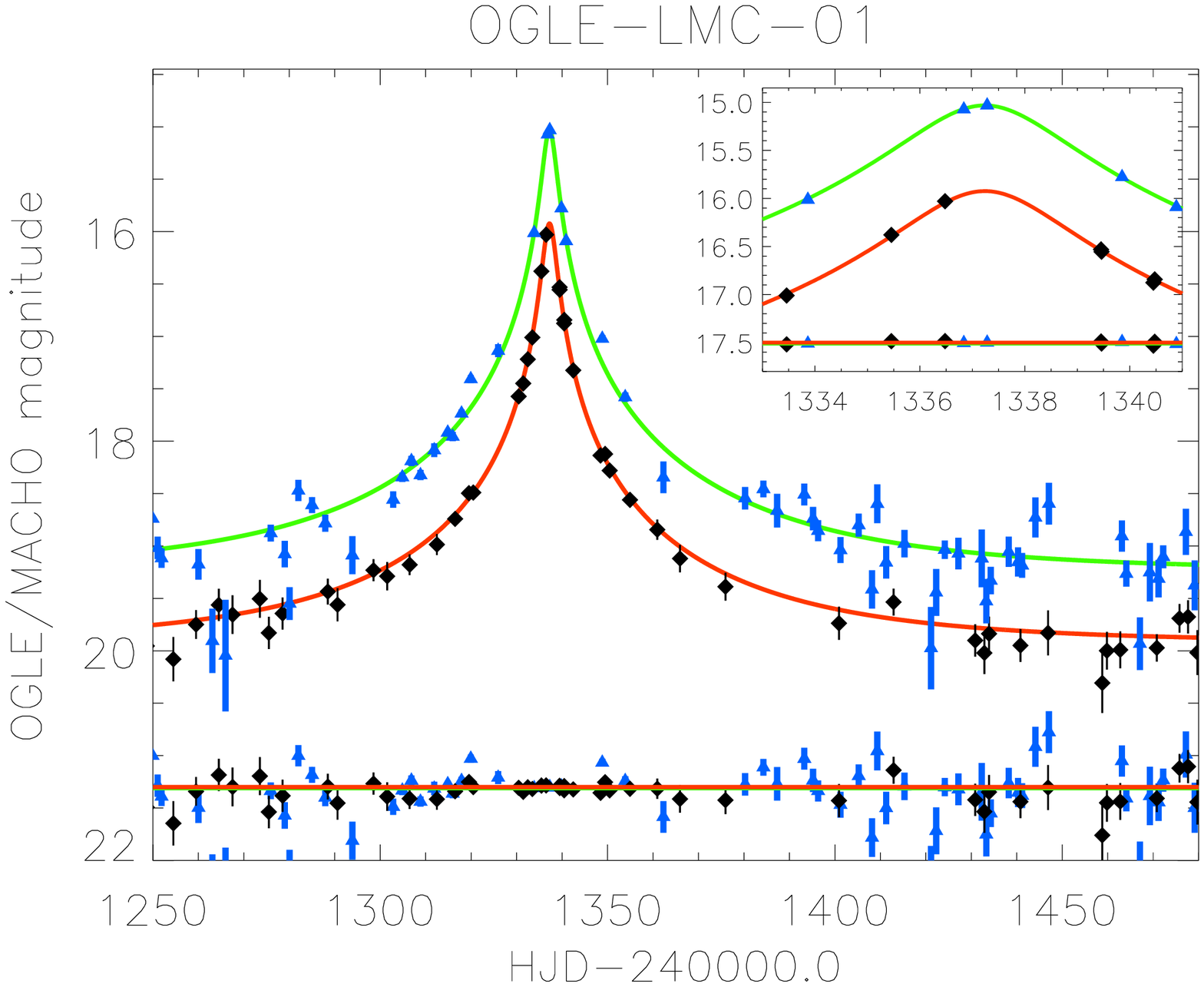}
\includegraphics[width=8.5cm]{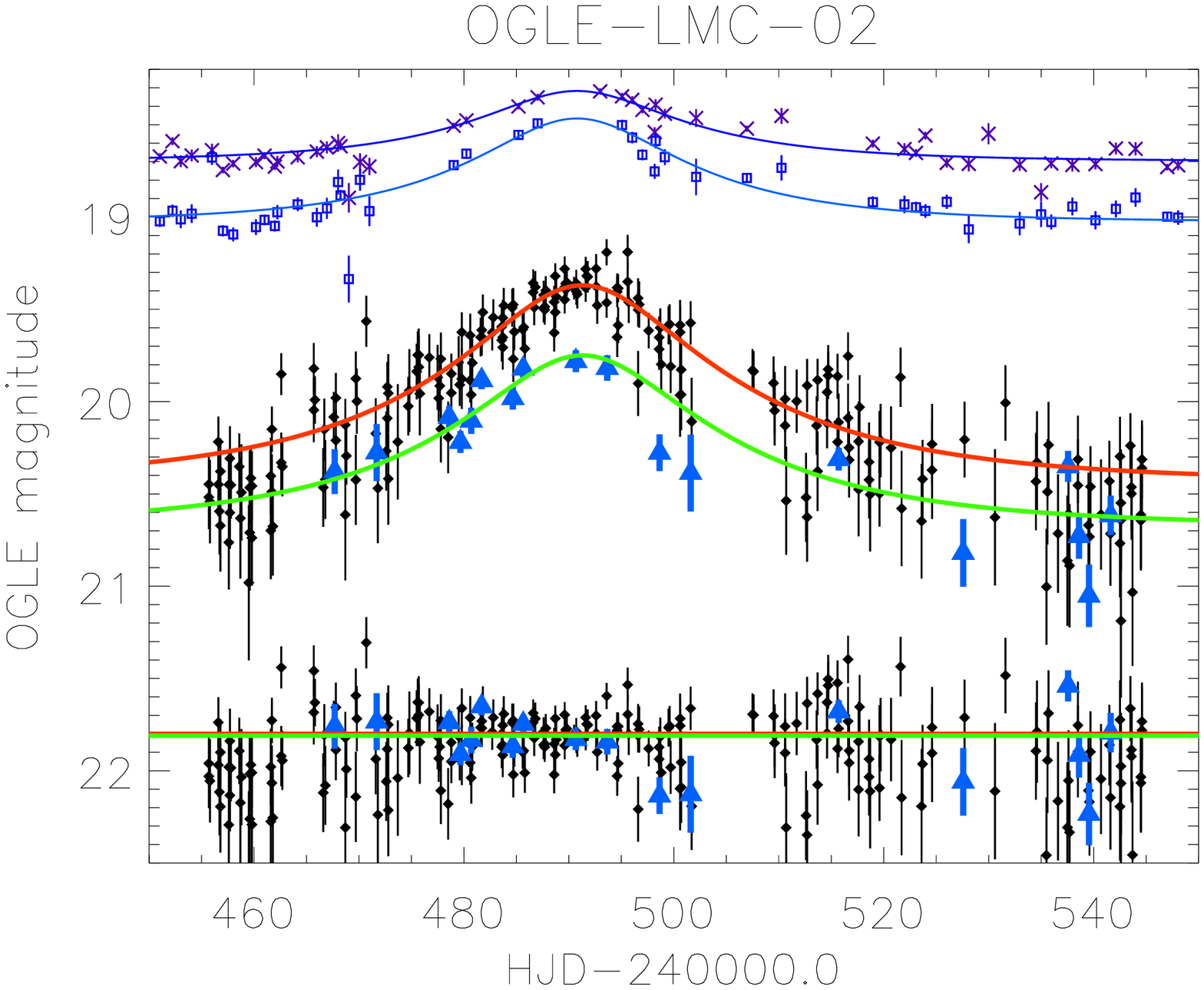}
\caption{Light curves of the two OGLE--II events OGLE-LMC-01 (left
  panel) and OGLE-LMC-02 (right panel).  Black points are the OGLE
  $I$ band, blue triangles are the MACHO $B$-band and OGLE $V$-band
  observations, for OGLE-LMC-01 and OGLE-LMC-02, respectively. Solid
  red and green lines show the best-fitting microlensing models to the
  data, with residuals shown below each curve. The inset shows a zoom-in around the peak of amplification,
  indicating that the finite-source effect is not detectable. Re-reduced
  MACHO data and the fit for OGLE-LMC-02 are shown above the OGLE data
  shifted by 0.5 mag (squares for B, crosses for R).  } 
\label{fig:events}
\end{figure*}

\section{Results}
\label{sec:results}

Out of $5.5 \times 10^6$ light curves we found only two plausible
candidates for microlensing events.  These are labeled OGLE-LMC-01 and OGLE-LMC-02.

Table \ref{tab:events} presents their coordinates, OGLE--II field,
star id, baseline $I-$ and $V$-band magnitudes obtained in the
microlensing fit and magnitude and colour of the source.

Both candidate light curves were checked for available additional data
outside the time-span of the OGLE--II project.  We cross-correlated them
with the OGLE--III database, covering years 2001--2008 (the
most-recent data used were obtained in May 2008) and the MACHO database, 
covering years 1992--1999, yielding total data coverage of 16 years.
Neither candidate exhibits additional variability apart from the
actual event.

In the OGLE-II database there were 44 $V$-band observations of
OGLE-LMC-01, but none of them were taken during the event. Therefore
for colour determination of the source, we used re-reduced MACHO data
in the $B$ band. By comparing the OGLE $V$ and $I$ CMD with MACHO $B$-band
photometry of nearby stars, we found the following transformation for
the colour: $(B-I)=(1.1571\pm0.0046)(V-I)+const$, where the measured constant is an artefact of the instrumental magnitude system of our reductions of the MACHO data.  
For OGLE-LMC-02, we used
the OGLE $V$-band data, as there were 59 observations available, with
about 14 taken during the peak.

Close-ups of the events are shown in Fig. \ref{fig:events} along with
the best-fitting microlensing models and their residuals.

\begin{table*}
\caption{ Parameters of the microlensing model fits to the OGLE-LMC-01 and OGLE-LMC-02 events. }
\label{tab:models}
\begin{tabular}[h]{lrrrrrr}
\hline
\multicolumn{7}{c}{OGLE-LMC-01}\\
parameter & \multicolumn{2}{c}{5-parameter fit} & \multicolumn{2}{c}{4-parameter fit} & \multicolumn{2}{c}{7-parameter fit}\\
\hline
$\t0$	\dotfill	& 1337.20  & $\pm0.02$	& 		1337.20	& $\pm0.02$ & 1337.20 & $\pm0.01$ \\
 & & & & & &\\
$\tE$	\dotfill	& 57.2	& $^{+4.7}_{-4.2}$		&  59.2	& $\pm0.7$ & 65.7 & $\pm2.6$ \\
 & & & & & &\\
 $\u0$	\dotfill	& 0.0258	& $^{+0.0024}_{-0.0023}$ & 0.0248 & $\pm0.0004$ & 0.0231 & $\pm0.0009$ \\
 & & & & & &\\
$\I0$  \dotfill	& 19.91 & $\pm0.01$ 		& 19.91 & $\pm0.01$ & 19.92 & $\pm0.01$ \\
 & & & & & &\\
 $\fsi$ \dotfill	& 1.04 & $\pm0.09$		&  1.0		& --- & 0.91 & $\pm0.04$ \\
 & & & & & &\\
 $B_0$  \dotfill	& --- 		& --- 		& --- 		& --- & 18.28 & $\pm0.01$ \\
 & & & & & &\\
 $f_{\rm S_B}$ \dotfill & --- & ---		&  ---		& --- & 1.11 & $\pm0.05$ \\
 & & & & & &\\
 $\chi^2$\dotfill  & 286.7 &  			 &   286.9 &   & 3401.0 &  \\
 & & & & & &\\
 ${\chi^2\over N_{dof}}$\dotfill  & 0.88	& 	&  0.87	&  & 2.68 &  \\
\hline
\hline
\multicolumn{7}{c}{OGLE-LMC-02}\\
parameter & \multicolumn{2}{c}{5-parameter fit} & \multicolumn{2}{c}{4-parameter fit} & \multicolumn{2}{c}{7-parameter fit}\\
\hline
$\t0$\dotfill		      & 491.6	& $\pm0.3$		& 491.6	& $\pm0.3$	& 491.3	& $\pm0.3$	\\
& & & & & & \\
 $\tE$\dotfill		      & 23.8 		& $^{+6.0}_{-5.1}$		&  24.2 		& $\pm1.0$	& 27.4	& $^{+6.9}_{-5.6}$\\
& & & & & &\\
 $\u0$\dotfill		      & 0.4120	& $^{+0.1856}_{-0.1164}$	& 0.4011	& $^{+0.0083}_{-0.0081}$	& 0.3478	& $^{+0.1428}_{-0.0955}$\\
& & & & & &\\
 $\I0$ \dotfill			& 20.42	& $\pm0.02$	&  20.42	& $\pm0.02$	& 20.44	& $\pm0.02$\\
& & & & & &\\
 $\fsi$\dotfill	      		& 1.04	& $^{+0.77}_{-0.37}$	&  1.0		& ---		& 0.83	& $^{+0.52}_{-0.28}$\\
& & & & & &\\
 $V_0$  \dotfill			& --- 		& --- 		& --- 		& --- & 	20.68	 & $\pm0.02$ \\
 & & & & & &\\
 $f_{\rm S_V}$ \dotfill 	& --- 		& ---		&  ---		& --- & 	0.67		& $^{+0.42}_{-0.22}$ \\
& & & & & &\\
 $\chi^2$\dotfill 			 & 578.2 	&  		&   578.2 &   						& 636.2	& \\
& & & & & &\\
 ${\chi^2\over N_{dof}}$\dotfill & 1.21 	& 		&  1.20 	& 					& 1.19	& \\
\hline
\end{tabular}

\end{table*}


The top panel of Fig. \ref{fig:cmd} shows the colour-magnitude
diagram of LMC stars based on OGLE and HST \citep{Holtzman2006} data, along with the determined positions of the sources in the 
two OGLE candidates.
For comparison, we have also included in this
figure the candidates from the MACHO collaboration:
different colours and shapes differentiate between
self--lensing candidates according to \citet{Mancini2004}, the binary event
MACHO-LMC-9, the thick disk lens candidates MACHO-LMC-5 \citep{MACHOLMC5}
and MACHO-LMC-20 \citep{MACHOLMC20} and the remaining candidates.  

In the bottom panel of Fig. \ref{fig:cmd}, the
positions of OGLE--II and MACHO candidates are shown on the map of
stellar density of the Red Clump stars from the OGLE--III data.

\subsection{OGLE-LMC-01}
This was the only event detected towards the LMC by the Early Warning
System (EWS, \citealt{Udalski1994ews}) during the entire second phase
of the OGLE project and was designated
OGLE-1999-LMC-01\footnote{http://www.astrouw.edu.pl/$\sim$ogle/ogle2/ews/1999/lmc-01.html}.
 It was also independently discovered and alerted by the MACHO
collaboration (MACHO-99-LMC-02), but it was not included in
\citet{AlcockMACHOLMC} as it occurred outside the 5.7 year time-span
of analysed data (MACHO star ID: 79.5863.4522,  see also \citealt{Bond2002}).

The OGLE--II data were fitted with the standard Paczy{\'n}ski
microlensing model, \ie a point-source--point-lens model (eq. 
\ref{eq:I}).  When we fixed blending such that the event was unblended
($f_{\rm S\mu 4}=1$) we obtained $t_{\rm E\mu 4}=59.2 \pm 0.7$ days,
whilst when $\fs$ was free, we found $\te =57.2^{+4.7}_{-4.2}$ days
($\fs=1.04 \pm 0.09$).  When OGLE $I$- and MACHO $B$-band data were
fitted simultaneously, the derived timescale was somewhat larger
($t_{\rm E_{VI}}=65.7$ days), as the blending parameter in $I$-band
changed slightly.  The results of the fits are summarised in Table
\ref{tab:models}.  

Since the event was relatively long we checked if OGLE-LMC-01
was subject to the parallax effect (\eg \citealt{HanGould1995MACHO}).  The components
of the parallax vector ${\mathbf \pi_{\rm E}}$ in the OGLE--II data
fit were $\pi_{{\rm E}, N} = -0.03 \pm 0.10$ and $\pi_{{\rm E}, E} =
-0.16 \pm 0.25$, \ie consistent at 1-$\sigma$ level with no detection
of parallax\citep{Gould2004LMC}.  Therefore, the
time-scale derived in the fit is not subject to any degeneracy with
parallax. In principle the lack of parallax signatures allows us to
place limits on lens distance. However, for this event the parallax
constraints are not very tight and so the limits are not
particularly insightful: if we assume that the source lies at the
distance of the LMC ($\sim50 \kpc$) and the lens is $0.5 \msun$, we
find that the 2-sigma lower-limit on the lens distance is 
300 pc. From this, we cannot make any significant statements about
the population to which the lens belongs.

With a maximum amplification of about 40 ($\sim 4$ mag), this is one
of the most secure and best constrained events towards the LMC found
to date.
High amplification events are often prone to exhibit finite-source effects in cases where the source and the lens are close to each other. We checked for the presence of this effect in the OGLE and MACHO data. In particular, MACHO data had two observations very close to the peak, where the effect should be most pronounced. We have not detected any deviation (see inset in Fig. \ref{fig:events}). This only allows us to put an upper limit on the source star size in units of the Einstein radii of $\rho > 0.0332$, which along with an estimation of the source size ($\theta_* = 0.42~{\rm \mu as}$) lead to an upper limit on $\theta_E = \theta_*/\rho > 13.1~{\rm \mu as}$. This at the LMC distance is equivalent to about $0.66$ AU. Hence, the limit on the projected transverse velocity of the lensing object $v_\bot=\theta_E/\te > 21$ km/s, which is satisfied by typical lenses in both the LMC and Galaxy halo.

A simultaneous microlensing model fit to the OGLE $I$-band and MACHO $B$-band data gives an apparent source magnitude of $I_s = 19.90 \pm0.06$ and colour $(V-I)_s = 0.887 \pm0.007$. Magnitudes and location of the source on the CMD in Fig. \ref{fig:cmd} indicate it belongs to the Red Giants Branch of the LMC.


\subsection{OGLE-LMC-02}

The second event, OGLE-LMC-02, was found just at the limit of our
magnitude cut (the actual mean brightness was 20.29 mag, which means
that it passed cut 0).  The maximum amplification of $\sim$2.6 produced a
rise in brightness of $\sim$1~mag.  This event occurred in the
very first season of the LMC observations in the OGLE--II, when the
sampling was much denser than in subsequent seasons.  Therefore, the
microlensing light curve was very well covered and prominent (see
Fig. \ref{fig:events}).

Again, we used the standard point-source--point-lens model to fit
the OGLE--II data.  The unblended fit ($f_{\rm S\mu 4}=1$) gave a
time-scale $t_{\rm E\mu 4}=24.2 \pm 1.0$ days, and for unconstrained 
$\fs$ we found $\te=23.8^{+6.0}_{-5.1}$ days
($\fs=1.04^{+0.77}_{-0.37}$). 
The large error bars in $\fs$ and $\te$ were also present in the simultaneous fit to $I$- and $V$-band observations, which yielded a time scale of $t_{\rm E_{IV}}=27.4$ days.

This object was also observed by the MACHO collaboration (MACHO star ID:
77.8152.2917). However, the event was not detected by them, probably due to noisy data and sparse sampling.
The peak was clearly revealed in both red and blue filters only after DIA re-reductions of the original MACHO images and its best-fitting microlensing model had a time-scale of around 25 days, consistent with the fit to the OGLE data. 

\begin{figure}
\center
\includegraphics[width=8.5cm]{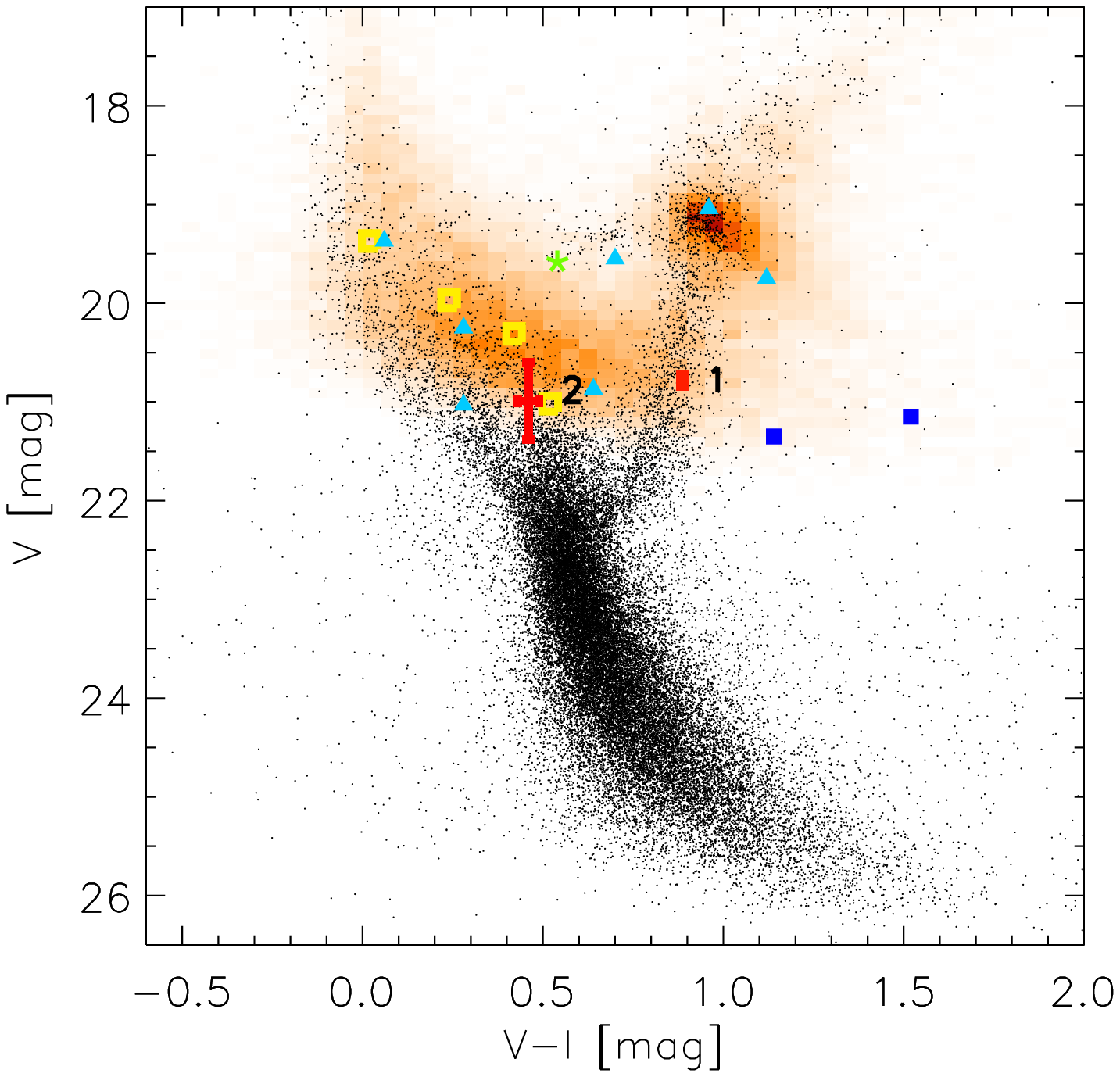}\\
\includegraphics[width=8.5cm]{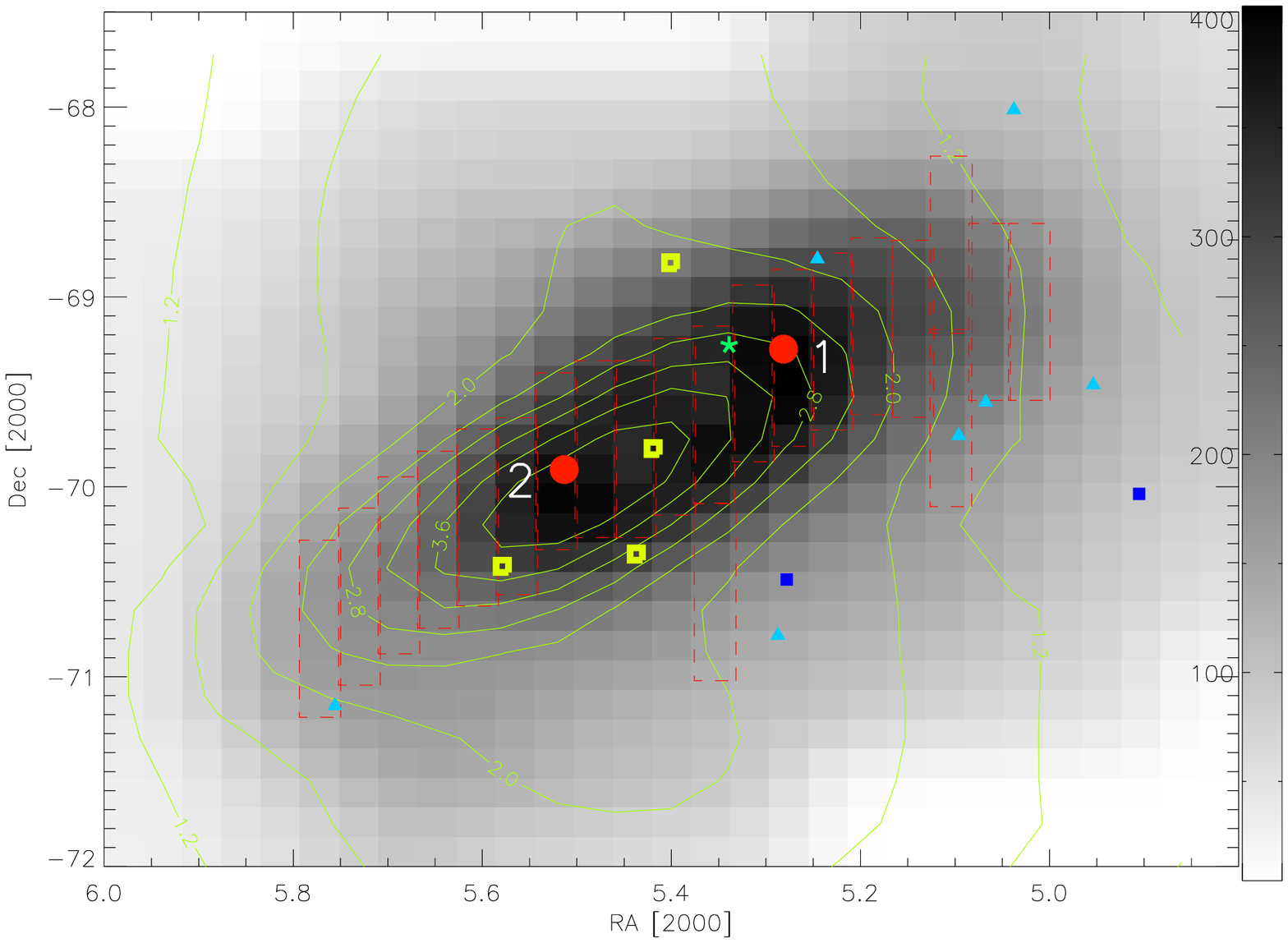}
\caption{  CMD of OGLE--II and MACHO candidate microlensing
events, overploted on the OGLE (red background) and HST (black dots)
measurements of stars in central parts of the LMC (upper panel).  
 Positions of OGLE and MACHO candidate events on the OGLE--III LMC 
Red Clump star count map (lower panel). 
The contours show modelled self-lensing optical depth from \citet{Mancini2004} and run every $0.4\times10^{-8}$ with $4\times10^{-8}$ at the innermost one. Dashed boxes mark OGLE--II fields.
Positions of the sources in the two OGLE events are marked in red and with a
number. Remaining symbols represent MACHO candidates with binary
event \#9 (green star), candidates for self--lensing (yellow
squares), confirmed thick-disk lenses \#5 and \#20 (dark blue filled
squares) and remaining marked with blue triangles.}
\label{fig:cmd}
\end{figure}

From the fit to the two-band OGLE data, we derived the magnitude and colour of the lensed source. The position of the source with magnitude $I_S=20.53\pm0.38$ and colour $(V-I)_S=0.46\pm0.03$, is marked on the Fig. \ref{fig:cmd} and indicates it is probably a Main Sequence star located in the LMC.

Its location on the Main Sequence branch of the CMD means that this
may potentially be a candidate for a blue bumper.  However, while many
blue bumpers exhibit secondary brightening episodes, the baseline of
this event proved to be constant over more than 15 years. Therefore, we
believe that it is unlikely to be a blue bumper.

\section{Blending and number of monitored stars}
\label{sec:blending}

Microlensing collaborations have deliberately chosen to monitor the
densest fields on the sky, namely the Galactic Center and Magellanic
Clouds. This is done in order to increase the number of potentially detectable
microlensing events, as the chances of seeing an events are of the
order of $10^{-7}$ (\ie one out of 10 million monitored stars is
undergoing a microlensing episode at any given time). However, in such
crowded stellar fields observed with a 
medium-sized ground-based telescope, virtually every object
detected on the image consists of several stars blended
together.  Neglecting blending can have severe consequences on
the optical depth measurements (\eg \citealt{SumiOGLEbulge}), as the
amplifications and time-scales of microlensing events may be
determined incorrectly, sometimes by a factor of 2 or more.  On the
other hand, \citet{Smith2007blending} showed that in the case of
the Galactic bulge the effect of overestimation of the events'
time-scales cancels out with the underestimate of the real number of
monitored stars to lowest order.

Dealing with blending is a delicate and difficult task.  Adding a
blending parameter to the standard microlensing light curve model may
cause degeneracies in the parameter space of the microlensing events
(\eg \citealt{Wozniakblending}).  Therefore, the favoured
approach in recent optical depth determinations towards the
Galactic bulge and Magellanic Clouds is to neglect blending by
narrowing the star sample to the brightest stars only
(\citealt{PopowskiMACHObulge}, \citealt{TisserandEROSLMC}). However,
in the case of Magellanic Clouds the number of possible
events is very small and it is even more unlikely to detect a
microlensing event occurring in the limited number of bright
stars. 

In our analysis, we decided to use all stars available in the OGLE--II
database.  To deal with the blending issue -- which is most serious
for faint stars -- we make use of the archival images from the Hubble
Space Telescope (HST)\footnote{http://archive.stsci.edu/hst/}
located by chance in our fields.  The HST images have superior
resolution when compared to the ground-based OGLE images (i.e. they are
nearly blending-free) and are usually much deeper than the OGLE data.
Therefore these data are a valueable source of information regarding
the amount of blending.
In the HST archive, we identified two deep ACS $I$-band
images coinciding with the OGLE fields, located in OGLE fields
LMC\_SC6 and LMC\_SC19, as typical representatives of the dense and sparse parts of the
LMC.  The location of these fields is shown in Fig. \ref{fig:fields}.

\begin{figure}
\center
\includegraphics[width=8.5cm]{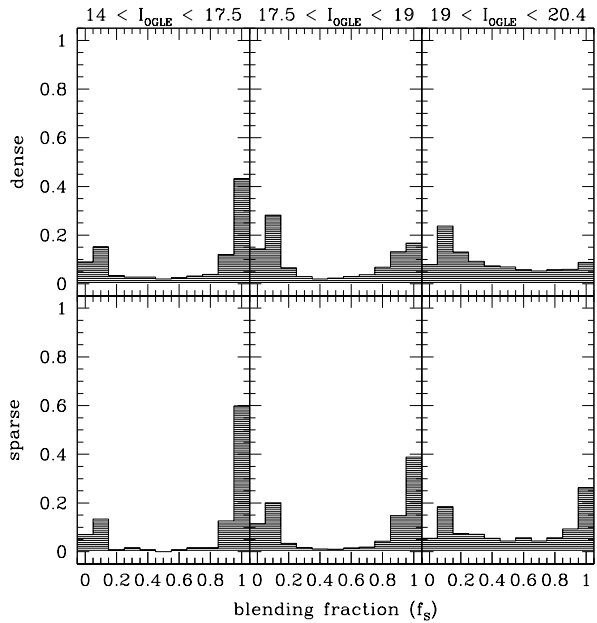}
\caption{Distributions of blending fractions for stars from the
  analysis of the archival high-resolution HST images of parts of the
  OGLE fields.}
\label{fig:fs}
\end{figure}

The positions of the HST stars were transformed onto the OGLE images
and then we identified which HST stars contributed to each OGLE object
by calculating their distances from the OGLE blend's centre.
A blend's radius was estimated using the Gaussian profile for the given
blend's magnitude at a chosen level above the background
($I\simeq19.7$ mag) with an additional 1.3 pixel added in quadrature.
We identified HST stars that were not more than 3.5~mag fainter than
the OGLE blend (\ie $\fs > 4$ per cent) and considered stars located within
the blend radius as components of that blend.  This method allowed us to create an
empirical distribution of the blending fractions for dense and sparse
fields.  We divided these distributions into three magnitude bins:
14~mag$<I\le$17.5~mag, 17.5~mag$<I\le$19~mag and 19~mag$<I
\le$20.4~mag. They are shown in Fig. \ref{fig:fs}.

Then, every OGLE LMC field was classified as dense or sparse according
to its average stellar density down to 20.4 mag (with 300 stars per square arcmin as a
division boundary) on the OGLE template image (see Table \ref{tab:fields}). We then
adopted the corresponding distributions in the detection efficiency
calculations and in our recalculation of the number of monitored stars
(see Section \ref{sec:eff}).

\begin{figure}
\center
\includegraphics[width=8.5cm]{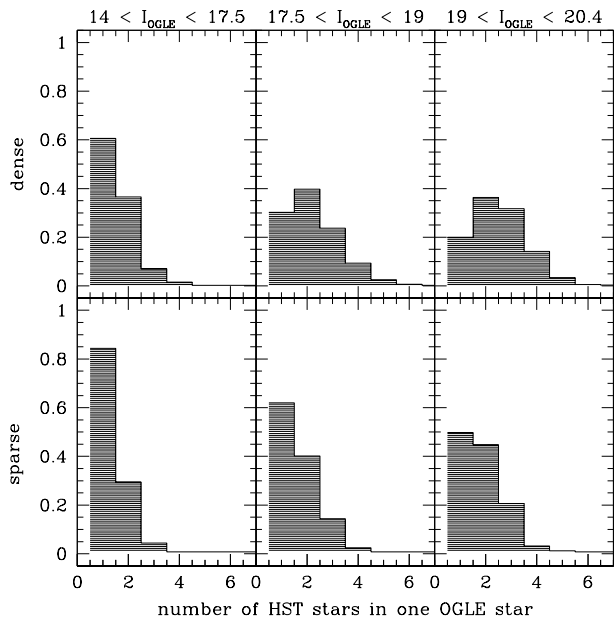}
\caption{Distribution of the number of HST stars in one OGLE--II object
  obtained for dense and sparse fields. }
\label{fig:nhst}
\end{figure}

Due to blending in ground-based observations, we can~not resolve
individual stars that are separated by less than the typical seeing (of
order of 1 arcsec).  Even though the OGLE template images, with
respect to the typical images, have usually high resolution thanks to
stacking of $\sim$a dozen of the best-seeing images, we still have
limited ability to resolve individual stars.  Since each OGLE template
object may consist of several stars, any of which could be microlensed,
the number of stars we consider as potential sources for microlensing
events is actually larger than number of objects detected on the
template image.

In order to estimate the actual number of monitored stars -- which
enters directly into eq. (\ref{eq:tau}) --
we derived the distributions of the number of HST stars hidden in a single
OGLE template object (see Fig. \ref{fig:nhst}) using the same HST
images and photometry. Then, for each OGLE field we convolved its
luminosity function with the distribution in each magnitude bin (for a
corresponding density level).  The estimated numbers of monitored
stars in each field are provided in Table \ref{tab:fields}. On
average, the correction factor for dense fields was about 2.3 and for
sparse about 1.7.  The total number of monitored stars was estimated
to be about 11.8 million, compared to about 5.5 million objects
detected on the OGLE--II template images in cut 0.

It is worth pointing out that in our method of counting the total
number of monitored stars we did not count all stars that
would be observed with the HST down to 23.9~mag (20.4 mag + 3.5 mag).
We counted only those stars which were blended with other stars such
that they formed a blend of magnitude brighter than 20.4~mag on
the OGLE template image. For instance, a single 21~mag star might not
be detected on the template image. Even 
if microlensed, we would not detect this event as in our analysis we
did not search for transient events with no baseline (\ie with no
object on the template).  If such star, however, was blended with
another 21~mag star they would form a 20.2~mag object -- very likely
detectable on the template. In this case, both components would have
blending factor of ${\fs}=0.5$ and both of them would be counted 
as a separate potential source for microlensing events.

\section{Detection Efficiency}
\label{sec:eff}

The task of calculating the detection efficiency for microlensing
events can be a complicated procedure. In general, this requires us to
simulate events and then determine the fraction that we recover
with our selection criteria. It may seem that the most accurate
method is to inject microlensing events into the series of real frames
or to create completely new mock images and then put them through the
entire photometric pipeline. However, this is obviously very
intensive computationally. Furthermore, it is not clear how to deal
with the fact that any star within the blended object can be
microlensed or whether the mock frame is indeed reproducing all
characteristics of the real one.  

Some of these problems can be also encountered when simulating microlensing
events at the so-called `catalogue-level' (which only involves generating series of light curves). However, the computational effort required for this procedure
is significantly reduced and some difficulties can be resolved more easily.

 The `image-level' simulations were performed by MACHO and MOA
collaborations in their calculation of the optical depth towards the
Galactic bulge \citep{AlcockMACHObulge, SumiMOAbulge, MACHOefficiency}.  
The `catalogue-level' approach has been more
commonly used for the bulge (\eg \citealt{AfonsoBulgeTau,
HamadacheEROSbulge, SumiOGLEbulge}) and in the EROS collaboration's study of
microlensing towards the Magellanic Clouds \citep{TisserandEROSLMC}.

In this study, we decided to use the existing photometric light
curves as a base for the simulations. However, as we discuss below, we
have attempted to incorporate all of the advantages of the
`image-level' simulations.  As shown in \citet{KozlowskiPhD} such an approach can be very
successful when all factors are carefully taken into account.

The detection efficiency was determined individually for the fields
containing both events (SC2 and SC8).
For a wide range of time-scales between 1 and 1000 days, we performed Monte Carlo simulation of about $10^5$ events. 
From the field's photometric database, a star was picked randomly,
provided that the star satisfied the basic cut 0 (\ie it had the
required number of data points and sufficient mean brightness).
For a given $\te$, the other microlensing parameters were drawn from
flat distributions ($\t0$ and $\u0$) and from the empirically derived
distributions for the blending parameter ($\fs$; see Section
\ref{sec:blending}).  The latter was chosen with respect to the
stellar density level of the simulated field and the magnitude bin in
which the simulated star was located.  The original flux of the input
star was apportioned between the source and blend in the ratio $\fs:1-\fs$.

Effectively, the lensed flux was added to already existing stars in
the database, so no new stars were `created'.  Also, the procedure
preserved any variability and non-gaussianity present in the original
photometry, allowing for effects of decreasing or increasing of the variability amplitude during microlensing \citep{varbaseline}.  Photometric errors-bars were scaled according to the
empirically derived relation between the error bars and the magnitude (see eq. \ref{eq:simerr}).
Such simulated events were then analysed with our search pipeline described in Section \ref{sec:search}.

\begin{figure}
\begin{center}
\includegraphics[width=8.5cm]{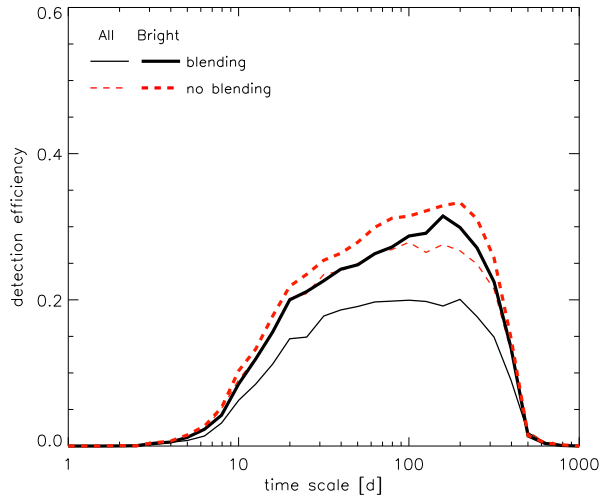}
\caption{Detection efficiencies for OGLE--II LMC\_SC8 field for All
  (thin lines) and  Bright (thick lines) Stars Samples, with blending included
  (solid lines) and neglected (dashed lines).  For
  most of the dense fields, the detection efficiencies follow
  these relations. The efficiencies for the most of the sparse fields are
  lower by a about 30 per cent due to the fact they have a shorter
  time-span of observations.
}
\label{fig:eff}
\end{center}
\end{figure}

\begin{figure}
\begin{center}
\includegraphics[width=8.5cm]{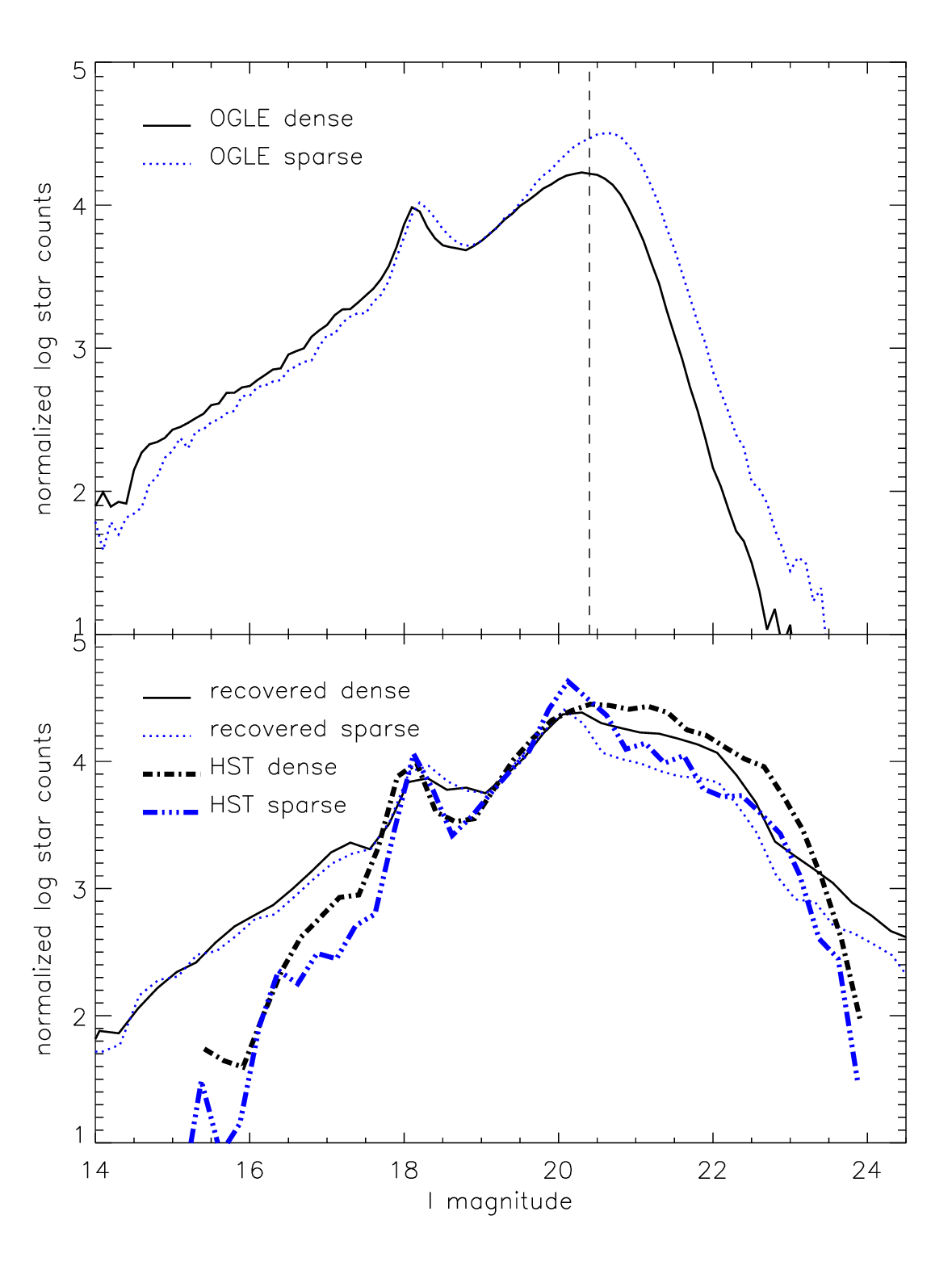}
\caption{Upper panel: observed luminosity functions in two
  OGLE--II LMC fields, where the dotted and solid lines represent
  sparse (SC19) and dense (SC6) fields, respectively. The vertical
  line shows the cut-off at $I=20.4$ mag.
  Lower panel: luminosity functions for the same fields recovered
  after applying the blending correction (see Section \ref{sec:blending}).
  Their shapes follow the prototype LFs of two HST fields (dense and
  sparse) used for the blending correction (thick dash-dotted lines).  }
\label{fig:lfs}
\end{center}
\end{figure}

In their LMC data, the MACHO collaboration found one event
(MACHO-LMC-9) to be a clear caustic crossing binary lens.  Our
detection pipeline discriminated against binary and other exotic events, as we
only fitted a point-source--point-lens microlensing model.  However,
we performed a visual inspection of the several thousands of high signal-to-noise light
curves that passed cut 1, and found no candidates for any
non-standard type of events.  Our detection efficiency did not take
into account the influence of binary lenses because Monte Carlo
simulations of binary events and their automated recovery is a
difficult and computationally demanding task. Therefore we 
followed the approach adopted by the EROS group \citep{TisserandEROSLMC} and 
reduced the efficiency by a factor of 0.9 to compensate for the 10 per cent possible binary lenses.

 We performed our simulations on both Bright and All Stars Samples
with blending included.  In addition, we repeated the analysis
assuming no blending ($\ie$ $fs=1$) in order to compare our results
with other studies.  We defined the Bright Stars Sample as stars
with $N>80$ and brighter than $I_{\rm RC} + 1$~mag, where $I_{\rm RC}$
is the Red Clump stars' centre on the colour-magnitude diagram.  In
the reference field LMC\_SC1, this threshold was at 18.8~mag.  The All
Stars Sample comprised of all stars used in the search for
microlensing events, \ie stars brighter than $20.4$ mag and with at
least 80 data points.

Fig. \ref{fig:eff} shows the derived detection efficiencies for a
wide range of time-scales for a typical dense field.  Most 
sparse fields were observed for a shorter period of time (observations
started about a year later than for dense fields).  However, in our
detection efficiency calculations we simulated microlensing events
within the full time-span of the OGLE--II experiment, namely for $\t0$
within the range of $HJD = 2450446 - 2451874$.  Therefore, the
efficiency for fields observed for a shorter time is reduced by about
30 per cent, owing to the events which fall outside the simulated light curve.
Only one field classified as sparse (LMC\_SC12) was monitored since
the very beginning of the project and the microlensing detection
efficiency for this field is comparable to the efficiencies
for dense fields.

Fig. \ref{fig:eff} also shows the detection efficiencies for the
non-blended analysis of All and Bright Stars Samples (with $\fs=1$ for all simulated events).  
The efficiency for Bright Stars with and without blending is higher than the 
corresponding efficiencies for All Stars Sample, as it is generally
easier to recover events occurring on brighter stars.

The observed luminosity functions (LFs) of dense and sparse OGLE--II
LMC fields are shown in the upper panel of Fig. \ref{fig:lfs}. 
Blending is more severe in dense fields and it causes faint stars
to `merge' and form brighter objects, populating brighter parts of the luminosity functions.  
The bottom panel of Fig. \ref{fig:lfs} shows
blending-free luminosity functions for a dense and sparse field, recovered using
the blending distributions obtained by comparing OGLE and
high-resolution HST images. The luminosity functions obtained from HST
are shown as well.  It is noticeable that the effect of
over-populating the brighter parts and under-populating the faint end
in blending-free LFs of dense fields has virtually disappeared after applying
the correction.  The recovered dense and sparse luminosity functions
follow their prototypes from the HST everywhere except at the bright and very faint end.
For stars brighter than $I\leq17$ mag the HST images become saturated, which is
reflected in the steeper luminosity functions at the bright end.  The
large scatter in the bright range is due to low number statistics in
the HST's small field-of-view (as compared to OGLE--II fields).
The broadening of the Red Clump may also be caused by this effect, or
could be due to the distance gradient along the OGLE field.
The HST's LFs of dense and sparse areas differ significantly, which is
likely caused by the different mix of LMC bar and disk populations in
these two fields.

In our Monte Carlo simulations, we drew an object from the original
OGLE luminosity functions.  Therefore we were not affected by the
incompleteness of the bright end of the HST luminosity
functions. Incompleteness in the faint end of the OGLE luminosity
functions is partially recovered in the correction process, but also
is negligible as the sensitivity for such faint stars is very low.
Although our blending analysis successfully recovers the underlying
stellar brightness distribution, it is still approximate as
we used only two HST images to represent dense and sparse fields. In
future, it would be interesting to expand this study with a
detailed $I-$band high-resolution follow-up, using HST and/or other
high-resolution instruments.

\section{Optical depth estimate}

We determined the optical depth using the All Stars Sample. 
For the two events found in the OGLE--II data we used the following standard
equation for calculating the optical depth:

\begin{equation}
\label{eq:tau}
\tau={{\pi}\over {2 N_* T_{\rm obs}}}\displaystyle \sum_i^{N_{\rm ev}} {{\tE}_{i} \over \epsilon({\tE}_i)}
\end{equation}
where $T_{\rm obs}=1428$ days is the time-span of all observations,
$N_*=11.8\times10^6$ is the total number of monitored stars (see Section
\ref{sec:blending}), $N_{\rm ev}=2$ is the total number of events, ${\tE}_i$
is the time-scale of each event detected with the efficiency of
$\epsilon({\tE}_i)$.  For the time-scales of these events we used the
microlensing fit where blending was a free
parameter (5-parameters model).  The error in $\tau$ was calculated with the formula
given by \citet{HanGouldtau}.  For the original detection efficiency (\ie
not corrected for binary events), the optical depth was derived to be
$\tau_{\rm LMC-O2} = 0.38\pm0.29\times10^{-7}$.  If the efficiency is
corrected for non--detectability of binary lenses, the optical depth
becomes $\tau_{\rm LMC-O2} = 0.43\pm0.33\times10^{-7}$.
Table \ref{tab:tau} presents all calculations of the optical depth.

\begin{table}
\caption{The optical depth for the two OGLE--II events.}
\label{tab:tau}
\begin{center}
\begin{tabular}[h]{cccc}
\hline
event & $\te$ & $\epsilon(\te)$ &$\tau_i \times 10^{-7}$ \\
\hline
\multicolumn{4}{c}{efficiency not corrected for binary events} \\
 & & & \\
OGLE-LMC-01 & $57.2^{+4.7}_{-4.2}$ & 0.212174 & 0.25 \\
 & & & \\
OGLE-LMC-02 & $23.8^{+6.0}_{-5.1}$ & 0.165536 & 0.13 \\
\hline
total $\tau_{\rm LMC-O2}$   &              &          &   $0.38\pm0.29$ \\
\hline
\hline
\multicolumn{4}{c}{efficiency corrected for binary events} \\
 & & & \\
OGLE-LMC-01 & $57.2^{+4.7}_{-4.2}$ & 0.190957 & 0.28 \\
 & & & \\
OGLE-LMC-02 & $23.8^{+6.0}_{-5.1}$ & 0.148982 & 0.15 \\
\hline
total $\tau_{\rm LMC-O2}$   &              &          &   $0.43\pm0.33$ \\
\hline
\end{tabular}
\end{center}
\end{table}


\section{Discussion}

\subsection{Optical depth}
The OGLE--II LMC data analysed in this study covered about 4.5 sq. deg.
of the sky over 4 years, which is clearly smaller than the data used
by MACHO or EROS groups.  However, it provides a completely independent data
set to test the hypotheses concerning the presence of MACHOs in the
Galactic halo.

Based on two the events we found the optical depth towards the LMC to
be $\tau_{\rm LMC-O2}=0.43\pm0.33\times 10^{-7}$.  On one hand, this
value is in 2-$\sigma$ agreement with the $\tau_{\rm LMC}=1.0\pm0.3$
obtained by MACHO \citep{BennettMACHOLMC}\footnote{Only 5 (\#1, \#8,
  \#14, \#15 and \#18) out of 17 MACHO candidates occurred within
  OGLE--II fields. However, all of these events peaked before the first
  observations of OGLE--II. Therefore we can only confirm continuity of their constant
  baselines using our data.  }.  
  If both OGLE--II events were caused by Machos, then they would contribute $8 \pm 6$ per cent to the total mass of the halo, 
  according to model {\bf {\it S}} from \citet{AlcockMACHOLMC}.

On the other hand, our value of
the optical depth is more consistent with a scenario of lensing due to
the visible stellar LMC component alone (self--lensing), for which various
estimates of the optical depth are present in the literature:
$\tau_{\rm SL}=(0.05-0.78)\times 10^{-7}$ \citep{Gyuk2000}, $\tau_{\rm
  SL}=(0.24-0.36)\times 10^{-7}$ \citep{AlcockMACHOLMC}, $\tau_{\rm
  SL}=0.54\times 10^{-7}$ \citep{Belokurov2004}.

\citet{Mancini2004} presented a detailed theoretical study of the halo and self-lensing optical depth and compared it with the results of MACHO.
They simulated the disk and bar of the LMC and quantitatively derived a contribution from the LMC's own stars to the optical depth as $\tau < 6\times10^{-8}$.
In the region of the LMC bar, where most OGLE-II fields are concentrated, they estimated the self-lensing optical depth to vary between $2$ and $4\times10^{-8}$.
The optical depth derived for two events we found in the OGLE-II data and averaged over all fields is in a perfect agreement with these estimates.

Also the time-scales of our two events seem to belong to the regime of
self--lensing, especially the 57.2-days event, which lies exactly on
the line for the self--lensing events in fig. 10 of
\citet{Mancini2004}.   From the CMD of the LMC, it is apparent
that both microlensed sources are coincident with an LMC population,
indicating that they probably belong to the LMC.   
The locations of both events in the bar of the LMC also supports the self--lensing scenario.

If we assume that our two events are indeed due to self--lensing
(\ie we have zero events caused by the halo objects), we can derive an
estimate for the upper limit of the Macho abundance in the halo of the
Milky Way.  We performed this analysis following the procedure similar to the one used by
EROS in \citet{TisserandEROSLMC}, first estimating the number of
expected events in the OGLE--II data set if the halo was entirely made of
Machos, using model {\bf {\it S}} \citep{AlcockMACHOLMC}.  In this
model the optical depth towards the LMC equals $\tau_{\rm
  total}^S=5.1\times 10^{-7}$, which also includes a
contribution from self--lensing and Galaxy disk lensing.  Dark matter
lensing solely had $\tau_{\rm Macho}^S=4.7\times 10^{-7}$.  For a so-called `typical' mass of $0.4~\msun$, in OGLE--II we should expect 16 events. 
At the 95 per cent Poisson confidence level there are up to 3 events consistent with a zero detection. This divided by a number of expected events gives the upper limit on the mass fraction of about 19 per cent.  
This translates to an upper limit on the optical depth due to Machos  
of $\tau_{0.4\msun} < 0.9\times 10^{-7}$.
If blending is neglected in the All Stars Sample the numbers are somewhat larger: about 9 expected events would put a limit
of $\tau_{0.4\msun} < 1.6\times 10^{-7}$.

\begin{figure}
\begin{center}
\includegraphics[width=8.5cm]{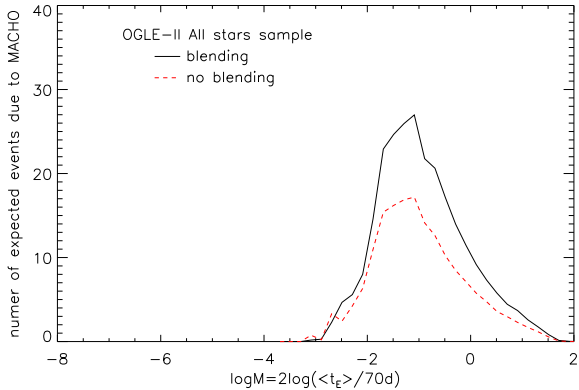}
\caption{The number of microlensing events caused by Machos expected to be
  seen in the OGLE--II data, for blended and non-blended approaches.}
\label{fig:nexp}
\end{center}
\end{figure}

\begin{figure}
\begin{center}
\includegraphics[width=8.5cm]{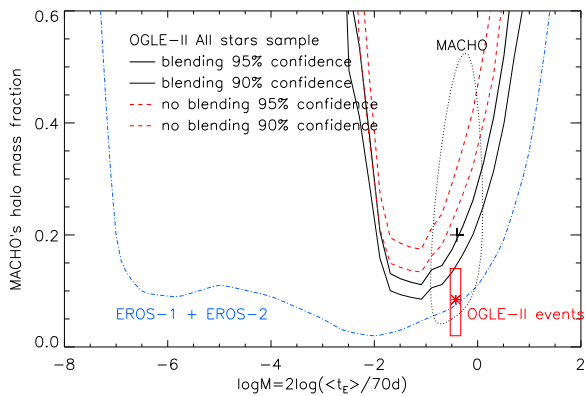}
\caption{  Mass fraction in compact dark halo objects as a function of the mass of
the lensing objects. The red box with a star shows the value for the two
OGLE--II events assuming they are caused by halo lenses.  Solid and
dashed lines show OGLE--II upper limits assuming both OGLE events are
due to self--lensing.  Also shown are the results of the MACHO
collaboration (dotted line) and the upper limit derived by the EROS
group (dot-dashed line). }
\label{fig:upperlimit}
\end{center}
\end{figure}

Fig. \ref{fig:nexp} shows the dependence of the number of expected events on
the mass of the lensing objects for the blended and non-blended approach.
Fig. \ref{fig:upperlimit} shows the 95 and 90 per cent confidence limits
on the halo Macho mass fraction for a wide range of masses. It also compares our results with the results from EROS and MACHO.  
This figure clearly shows that the OGLE--II project is the most
sensitive to events caused by Machos with masses between 0.01 and 0.2
$\msun$.  
For this range of masses, we can place an even tighter upper-limit of
about 10 per cent (at 95 per cent confidence) on the abundance of compact dark matter
objects in the Galactic halo. A more detailed theoretical analysis will be presented in a
forthcoming paper (Calchi Novati, Mancini, Scarpetta et al., in preparation).

For the sake of comparison with the results of the EROS group, we also
performed the above analysis narrowing the stars sample only to
the bright stars.  In each field, our Bright Sample consisted of about
20 -- 30 per cent of stars present in All Stars Sample, with a total over all
fields of about 1.9 million. After correction for blending, the number
of monitored stars was about 3.6 million.  In the Bright Sample of
OGLE--II, we have not found any events. According to the full $S$ model
of \citet{AlcockMACHOLMC}, we should expect about 9 events to be
caused by lenses of mass $0.4\msun$, including events due to
self--lensing. If we neglect blending, then we expect 6 events.
The total optical depth towards the LMC is therefore estimated to be
$\tau_{0.4\msun} < 1.6\times 10^{-7}$ ($\tau_{0.4\msun} < 2.6\times
10^{-7}$) with (without) the full blending correction.  The calculated upper-limit
is very high due to the fact that OGLE--II has much smaller sky
coverage and shorter time-baseline compared to the combined EROS--I
and EROS--II projects.  
However, a combined study of OGLE--II and OGLE--III data will  be
conducted in the near future and should be able to yield
constraints of even stronger magnitude than those given by the EROS collaboration
(\citealt{WyrzykowskiManchester2008}, Wyrzykowski et al., in
preparation).

\subsection{Blending}
As has been discussed above, we did not have HST data for all 
OGLE--II fields. Therefore 
we chose to use the two selected HST fields and consider these as being
representative of two different density levels.  In this way we
recover only an approximation to the underlying real luminosity
function for each field.

In order to check how our selection of HST stars affects the result, we
repeated the whole analysis for a different cut on the HST stars
magnitude, namely 2.5~mag (instead of 3.5~mag) below the brightness of
the OGLE objects. This translates to a minimal allowed blending
fraction of 10 per cent.  We then rederived the blending distributions and
the distributions for the number of HST stars in a single OGLE
object.  These distributions were obviously different compared to
the original 3.5 mag cut; for both sparse and dense fields, there
was nearly no blending in the brightest magnitude bin.  For the
fainter stars, the distributions indicated a level of blending similar
to that which was found previously, but with slightly less pronounced
features for small $\fs$.

In a similar manner to that described in Section \ref{sec:blending},
we estimated the total number of monitored stars and found about
$9.2\times10^6$ stars.  Then we also repeated the efficiency
calculation including these new blending distributions. It was similar
to the original efficiency to within 2 per cent.  The detection
efficiencies (corrected for binary events) for the two events were
found to be $\epsilon(\te=57.2)=0.214$ and $\epsilon(\te=23.8)=0.165$.
Since these efficiencies are similar to previous values while the
number of monitored stars is reduced, we find that this yields a
somewhat larger optical depth of $\tau_{\rm HST2.5}=(0.49\pm
0.37)\times10^{-7}$.  This is, however, still consistent with our
original value, indicating little sensitivity of the optical depth on
the depth of blending stars' magnitude limit.

As an extreme case we also considered neglecting blending completely
in our All Stars Sample. In this case, we find that there are only
about 5.5 million sources with $\fs=1$. The detection efficiency was
obtained as described above and is shown on Fig. \ref{fig:eff}.  It was larger
than in the case when blending is taken into account by about
20 per cent. However, this does not cancel out entirely with the reduction in
the number of monitored sources. Using these numbers with non-blended model
time-scales (Table \ref{tab:models}) leads to a larger value for
the optical depth of 
$\tau=(0.72\pm0.55)\times10^{-7}$, where we have used the detection
efficiencies corrected for lack of binary events.

This indicates that neglecting blending favours larger values of the optical
depth, emphasising the importance of carrying out a careful analysis
of blending when dealing with All Stars samples in order to obtain reliable optical depth results towards the LMC.

\section{Conclusions}

In the search for microlensing events in OGLE--II data towards the
Large Magellanic Cloud, we found two events.  The total optical depth
for all 21 OGLE--II LMC fields, covering about 4.5 square degrees, was
derived for the All Stars Sample (with a limit of $\langle I \rangle
\le 20.4$ mag) 
as $\tau_{\rm LMC-O2}=0.43\pm0.33~10^{-7}$.  
If both events were caused by Machos, this would imply that their comprise $8 \pm 6$ per cent of the halo by mass.

However, the value of the optical depth and the characteristics of the events are consistent with the
self--lensing scenario, in which lensing towards the LMC occurs only
between sources and lenses both located in the Cloud.  From this, we
can constrain the presence of compact dark matter objects in the Milky
Way's halo and we find that our results are compatible with no
detection. The upper limit is found to be 19 per cent for
$M_{Macho}=0.4 \msun$ and 10 per cent for masses between 0.01 and 0.2
$\msun$.  It is worth emphasising that in our study we probed
mainly the bar of the LMC, similarly to the MACHO project. 
However, our results and conclusions are closer to the ones of the EROS
collaboration, which probed different regions of the LMC.  

The sensitivity of the OGLE--II experiment is limited to a relatively
short time span and sky coverage. However, the results presented in
this paper will be soon strengthened and supplemented with the
analysis of the 8-year OGLE--III data.

Future photometric and astrometric surveys may confirm the lensing
origin of most of events detected so far and the new ones by high-resolution imaging or
precise astrometry, allowing for measuring the distances to the
sources and lenses or their velocities.

\section*{acknowledgements}
The authors would like pay tribute to the late Bohdan Paczy{\'n}ski,
as the subject of this work was always a subject of interest for him.
We are grateful to the referee, Andy Gould for his comments and
assistance in combining MACHO and OGLE data sets for measure the colour of OGLE-LMC-01.  
We also would like to thank for their help at various stages of this work to Drs Wyn
Evans, Subo Dong, Luigi Mancini, Sebastiano Calchi-Novati, Gaetano
Scarpetta, Hongsheng Zhao and Daniel Faria.  {\L}W, SK, MCS and JS
acknowledge generous support from the European Community's FR6 Marie
Curie Programme, Contract No. MRTN-CT-2004-505183 ``ANGLES''.  MCS
also acknowledges support from the STFC-funded ``Galaxy Formation and
Evolution'' program at the Institute of Astronomy, University of
Cambridge.  The OGLE project is partially supported by the Polish
MNiSW grant N20303032/4275.  JS also acknowledges support through the
Polish MNiSW grant no. N20300832/070.

\bibliographystyle{mn2e}

\label{lastpage}
\end{document}